\documentclass[iop,apj]{emulateapj}
\usepackage{amsmath}
\usepackage{rotating}

\def\stacksymbols #1#2#3#4{\def\theguybelow{#2}
        \def\verticalposition{\lower#3pt}
        \def\spacingwithinsymbol{\baselineskip0pt\lineskip#4pt}
        \mathrel{\mathpalette\intermediary#1}}
\def\intermediary #1#2{\verticalposition\vbox{\spacingwithinsymbol
        \everycr={}\tabskip0pt
        \halign{$\mathsurround0pt#1\hfil##\hfil$\crcr#2\crcr
                \theguybelow\crcr}}}

\shorttitle{Thermal Conduction in Galaxy Clusters}
\shortauthors{Fang et al.}

\begin{document}
\bibliographystyle{apj} 

\title {On the Efficiency of Thermal Conduction in Galaxy Clusters}

\author{Xiang-Er Fang$^{1}$, Fulai Guo$^{2,3}$, Ye-Fei Yuan$^{1}$, and Guobin Mou$^{1}$}

\affil{$^{1}$Key Laboratory for Research in Galaxies and Cosmology, Department of Astronomy, University of Science and Technology of China, Hefei, Anhui 230026, China}
\affil{$^{2}$Key Laboratory for Research in Galaxies and Cosmology, Shanghai Astronomical Observatory, Chinese Academy of Sciences,
80 Nandan Road, Shanghai 200030, China; fulai@shao.ac.cn}
\affil{$^{3}$School of Astronomy and Space Science, University of Chinese Academy of Sciences, 19A Yuquan Road, 100049, Beijing, China}

\begin{abstract}

Galaxy clusters host a large reservoir of diffuse plasma with radially-varying temperature profiles. The efficiency of thermal conduction in the intracluster medium (ICM) is complicated by the existence of turbulence and magnetic fields, and has received a lot of attention in the literature. Previous studies suggest that the magnetothermal instability developed in outer regions of galaxy clusters would drive magnetic field lines preferentially radial, resulting in efficient conduction along the radial direction. Using a series of spherically-symmetric simulations, here we investigate the impact of thermal conduction on the observed temperature distributions in outer regions of three massive clusters, and find that thermal conduction substantially modifies the ICM temperature profile. Within 3 Gyr, the gas temperature at a representative radius of $0.3r_{500}$ typically decreases by $\sim 10 - 20\%$ and the average temperature slope between $0.3r_{500}$ and $r_{500}$ drops by $\sim 30 - 40\%$, indicating that the observed ICM would not stay in a long-term equilibrium state in the presence of thermal conduction. However, X-ray observations show that the outer regions of massive clusters have remarkably similar radially-declining temperature profiles, suggesting that they should be quite stable. Our study thus suggests that the effective conductivity along the radial direction must be suppressed below the Spitzer value by a factor of $10$ or more, unless additional heating sources offset conductive cooling and maintain the observed temperature distributions. Our study provides a smoking-gun evidence for the suppression of parallel conduction along magnetic field lines in low-collisionality plasmas by kinetic mirror or whistler instabilities. 
 
\end{abstract}


\section{Introduction}
\label{section1}

As the largest gravitationally-bound systems in the Universe, galaxy clusters contain a large amount of hot, diffuse plasma, which is usually referred as the intracluster medium (ICM). X-ray observations by {\it Chandra} and {\it XMM-Newton} have been widely used to measure the density and temperature distributions of the ICM, finding that typical electron number densities of the ICM are in the range of $10^{-3}$ - $10^{-1}$~cm$^{-3}$ and typical temperatures are $1$ - $15$ keV (\citealt{PetersonFabian2006}). Presumably heated by gravitational infall, the ICM is ionized and weakly collisional with a typical Coulumb mean free path $\lambda \sim 3(T/10 \rm{~keV})^{2}(n_{\rm e}/0.01 \rm{~cm}^{-3})^{-1}$ kpc, where $T$ and $n_{\rm e}$ are the temperature and electron number density of the ICM, respectively. The ICM is also weakly magnetized with a typical magnetic field strength $B \sim 1-10$ $\mu$G, corresponding to a large plasma beta ($\beta=8\pi P/B^{2}$) of typically several hundreds (see \citealt{CarilliTaylor2002} and \citealt{Feretti2012} for recent reviews). The magnetic and low-collisionality nature of the ICM makes it to be an ideal laboratory to study plasma astrophysics. 

The temperature profiles of many galaxy clusters have been measured by X-ray observations to a concentric radius of nearly $r_{500}$ or beyond, and show remarkable similarity in the outer parts of galaxy clusters beyond $\sim 0.1r_{200}$ with radially-declining temperature profiles (\citealt{Vikhlinin2005}; \citealt{sanderson06}; \citealt{baldi07}; \citealt{pratt07}; \citealt{leccardi08}; \citealt{zhu16}). Here $r_{\Delta}$ is the radius enclosing a mean matter density of $\Delta$ times the critical density of the Universe at the cluster redshift. The observed temperature gradients in the high-temperature ICM point to the potential importance of thermal conduction on the evolution of galaxy clusters, where the conduction timescale based on the classic Spitzer conductivity \citep{Spitzer1962} in an unmagnitized plasma could be substantially shorter than the cluster age. Thermal conduction in the ICM has thus attracted a lot of attention in the literature, particularly for cool-core clusters where inward conduction into low-temperature cool-core regions may potentially be an important energy source to solve the so-called cooling flow problem (\citealt{NarayanMedvedev2001}; \citealt{ZakamskaNarayan2003}; \citealt{guo08}; \citealt{guo08b}). 

Due to the complications caused by magnetic fields and turbulence, the efficiency of thermal conduction in the ICM is a long-standing problem in astrophysics. Since the electron gyroradius  is many orders of magnitude smaller than its Coulumb mean free path, thermal conduction in the ICM is expected to be highly anisotropic and dominated by parallel conduction along magnetic field lines (e.g., \citealt{NarayanMedvedev2001}; \citealt{clark16}). The ICM is likely turbulent (see, e.g., \citealt{InogamovSunyaev2003}; \citealt{Schuecker2004}; \citealt{Zhuravleva2014}), and the tangled magnetic fields may strongly suppress effective thermal conduction by a factor of $10^2$ to $10^3$ (\citealt{ChandranCowley1998}; \citealt{RechesterRosenbluth1978}). However, if the turbulence extends over a wide range of length scales, \citet{NarayanMedvedev2001} argued that thermal conductivity can be boosted to be $\sim 1/5$ of the classic Spitzer value. Such efficient thermal conduction could contribute significantly to offsetting radiative cooling from cool core regions of massive clusters (\citealt{ZakamskaNarayan2003}; \citealt{guo08}; \citealt{guo08b}; \citealt{jacob17}). However, cool cluster cores typically exhibit radially increasing temperature profiles with positive temperature gradients, and are susceptible to the heat-flux-driven buoyancy instability (HBI; \citealt{quataert08}), which may saturate by rearranging the magnetic field lines to be perpendicular to the radial direction, resulting in strong suppression of radial thermal conduction (\citealt{parrish08b}; \citealt{mccourt11}).\footnote{Strictly speaking, this is only correct in the extremely high $\beta$ case (i.e. when the initial magnetic fields are very weak). For intermediate field strengths more realistic for real clusters, \citet{Kunz12} and \citet{Avara13} show that vertical magnetic filaments form and are HBI-stable, enabling significant conductive heat flux.}

On the other hand, the outer regions of clusters typically show radially declining temperature profiles with negative gradients, and are susceptible to the magnetothermal instability (MTI; \citealt{balbus00}), which would quickly reorient magnetic field lines to be preferentially radial (\citealt{ParrishStone2007}; \citealt{Parrish2008}; \citealt{Sharma2008}). Therefore, thermal conduction becomes very efficient along the radial direction, and is capable of substantially modifying the ICM temperature profiles in massive clusters within several billion years (\citealt{Parrish2008}). The outer regions of massive galaxy clusters are thus an ideal place to investigate the efficiency and impact of thermal conduction in diffuse, low-collisionality plasmas. 

More recently, theoretical studies and particle-in-cell simulations indicate that parallel thermal conduction along magnetic field lines in a low-collisionality high-$\beta$ plasma may be significantly suppressed by kinetic mirror or whistler instabilities on the length scale of the ion or electron gyroradius, respectively (\citealt{Komarov2016}; \citealt{Riquelme16}). These kinetic-scale instabilities are triggered by pressure anisotropies resulted from plasma motions and amplify local magnetic fields, which pitch-angle scatter electrons and suppress the transport of heat. \citet{clark16}, \citet{clark18}, and \citet{komarov17} further show that the electron heat flux could self-generate turbulence and drive whistler waves, which in turn strongly suppress thermal conduction. These studies suggest that thermal conduction in the ICM may be very inefficient. Nonetheless, observational evidences for the importance of these plasma microphysics processes are still scarce.

To investigate the efficiency of thermal conduction in galaxy clusters, in this paper we use a series of hydrodynamic simulations to study the impact of thermal conduction on the temperature profiles in the outer regions of three massive clusters: A1795, A2029, and A478. The temperature and density profiles of these clusters have been determined by {\it Chandra} X-ray observations out to a large radius $\sim r_{500}$ or beyond \citep{Vikhlinin2005}. The outer regions of these clusters show similar radially-declining temperature profiles, and if electron heat conduction along magnetic field lines is not suppressed by kinetic instabilities, the MTI would quickly develop there and reorient field lines to be preferentially radial. Thermal conduction in the radial direction is thus expected to be very efficient with an effective conductivity estimated to be close to half of the Spitzer value \citep{Parrish2008}. We specifically choose these three massive clusters with high gas temperatures, corresponding to relatively short Spitzer conduction timescales. Starting with the observed ICM temperature and density profiles in hydrostatic equilibrium, we investigate the evolution of the ICM temperature profile under the influence of various levels of thermal conduction, and aim at constraining the efficiency of thermal conduction with respect to the classic Spitzer value and finding smoking-gun signatures of kinetic-scale instabilities in the ICM.

The rest of the paper is organized as follows. In Section~\ref{section:method}, we describe the details of our methods, including hydrodynamic equations, modeling of thermal conduction,
the ICM model, initial conditions, and the simulation setup. Then, we present the simulation results on the ICM temperature evolution in Section~\ref{sec:Tevol}, and investigate the physics behind the ICM temperature evolution in Section \ref{sec:physics}. We discuss important implications of our simulations in Section \ref{sec:implications}. Finally, we summarize our main results in Section~\ref{section:summary}.

\section{Methods}
\label{section:method}

To the first-order approximation, the ICM distribution in galaxy clusters is spherically symmetric, which is also assumed in our model. X-ray observations indicate that, when scaled with the average cluster temperature and the virial radius, the radially-declining temperature profiles in outer regions ($r \gtrsim 0.1r_{200}$) of galaxy clusters show remarkable similarity (e.g., \citealt{Vikhlinin2005}; \citealt{pratt07}), suggesting that they are likely long-lasting and stable. This is roughly consistent with predictions from cosmological simulations where the physics of thermal conduction has not yet been included (e.g., \citealt{Loken2002}; \citealt{Borgani2004}; \citealt{hahn17}), and further suggests that the temperature profiles in outer regions of massive clusters may be used to constrain the efficiency of thermal conduction in the ICM. If thermal conduction along the radial direction is efficient, it may substantially modify the temperature profiles so that they become inconsistent with observations. Along with these thoughts, in this paper we use a series of hydrodynamic simulations to investigate the evolution of the radial temperature profiles of three massive clusters A1795, A478 and A2029, under the influence of various physically-motivated levels of thermal conduction. In this section below, we present the details of our methodology.

 \begin{figure*}
    \centering
 \plottwo{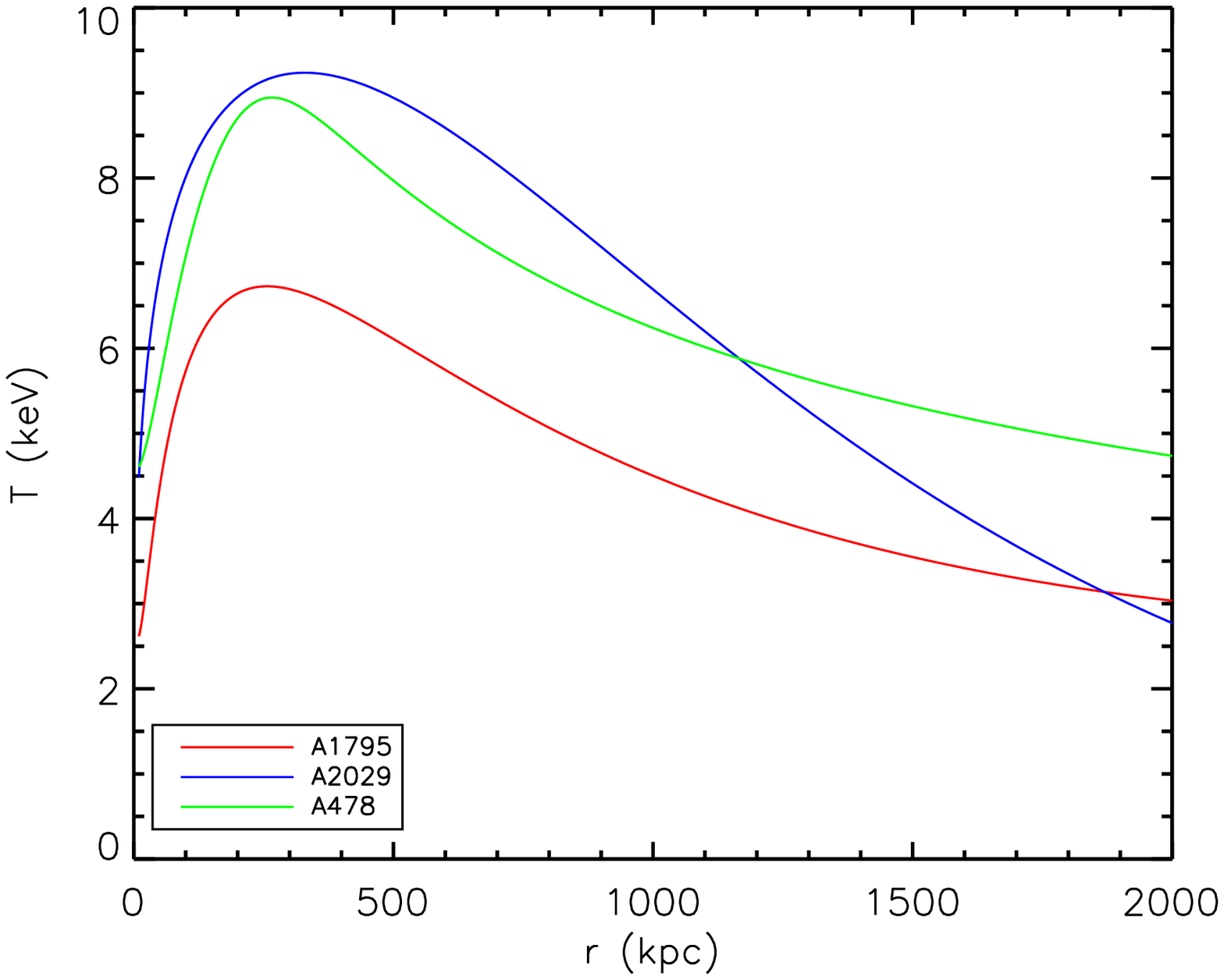}{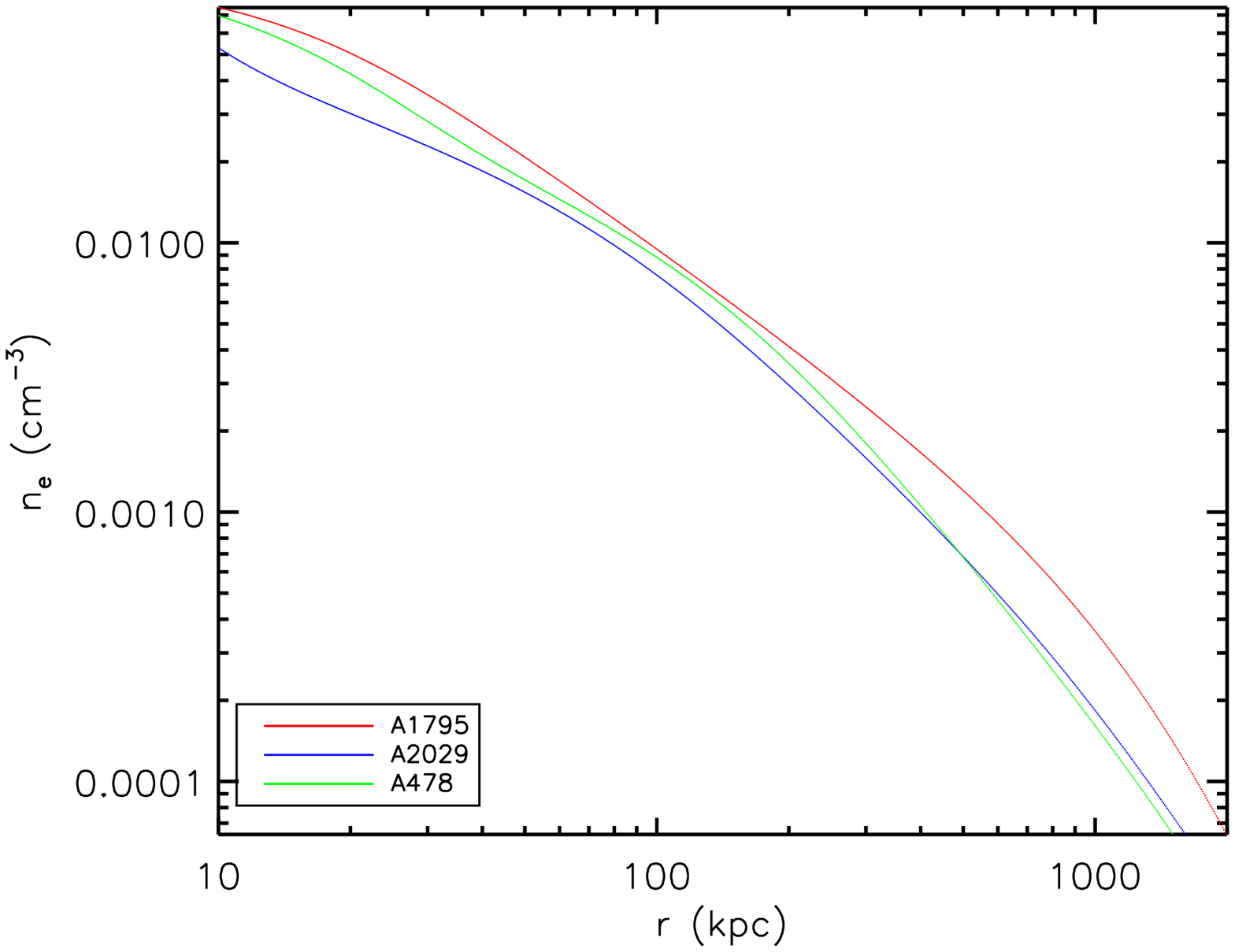}
 \caption{Initial ICM temperature ({\it left}) and electron number density ({\it right}) profiles in the galaxy clusters studied in the present paper: A1795 ({\it red}), A2029 ({\it blue}), and A478 ({\it green}). These analytic profiles are observational fits to {\it Chandra} data presented in \citet{Vikhlinin2006}.}
 \label{plot1}
 \end{figure*}

\subsection{Basic Equations}
\label{equations}
We approximate the weakly-collisional ICM as a thermal fluid, and study its evolution under the influence of gravity and thermal conduction. The basic hydrodynamic equations governing the ICM evolution may be written as:
\begin{eqnarray}
\frac{d \rho}{d t} + \rho \nabla \cdot {\bf v} = 0,\label{hydro1}
\end{eqnarray}
\begin{eqnarray}
\rho \frac{d {\bf v}}{d t} = -\nabla P-\rho \nabla \Phi ,\label{hydro2}
\end{eqnarray}
\begin{eqnarray}
\frac{\partial e}{\partial t} +\nabla \cdot(e{\bf v})=-P\nabla \cdot {\bf v}-\nabla \cdot {\boldsymbol{F}}- \mathcal{C},\label{hydro3}
   \end{eqnarray}
\noindent
where $\rho$, ${\bf v} $, $P$, $e$ are the density, velocity, pressure, and energy density of the ICM, respectively. $\Phi$ is the gravitational potential in the ICM (see Section~\ref{initcond}), and $d/dt \equiv \partial/\partial t+{\bf v} \cdot \nabla $ is the Lagrangian time derivative. $\mathcal{C}$ is the radiative cooling rate per unit volume, and as the cooling time in outer regions of galaxy clusters is typically longer than the Universe's age, it is set to $\mathcal{C}=0$ in our simulations (see Fig. \ref{timescales} and the relevant text in Sec. \ref{sec:implications}). These basic hydrodynamic equations are closed by the relation $P=(\gamma-1)e$, where $\gamma=5/3$ is the adiabatic index. 

The gas temperature $T$ and electron number density $n_{\rm e}$ can be derived from the ideal gas law:
\begin{equation} 
  \label{stateequ}
 P=\frac{\rho k_{B} T}{\mu m_{\mu}} =  \frac{\mu_{\rm e}}{\mu} n_{\rm e}k_{B} T {,}
  \end{equation}
 where $k_{B}$ is Boltzmann's constant, $m_{\mu}$ is the atomic mass unit, and $\mu = 0.62$ and $\mu_{\rm e} = 1.18$ are the mean molecular weight per particle and per electron, respectively, corresponding to a fully ionized plasma with the Helium mass fraction $Y=0.28$ (\citealt{ZakamskaNarayan2003}; \citealt{guo08}).

\begin{table}
	\centering
   \begin{minipage}{70mm}
   	\caption{List of Simulations}
   	\label{simulatable}
   	\begin{tabular}{@{}llcc}
   	 \hline
	  \hline
   	    Cluster & Run& ~~~~$f_{\rm in}$ ~~~&~~~$f_{\rm out}$\\
   	 \hline
   	 A1795........................& A0 &0&0\\
   	   & A1Y & 0.1 &0.1 \\
	   & A1N & 0 &0.1 \\	   
    	   & A3Y & 0.3 &0.3 \\
	   & A3N & 0 &0.3 \\	 
    	   & A5Y & 0.5 &0.5 \\
	   & A5N & 0 &0.5 \\	 
   	 A478.........................& B0 &0&0\\
    	   & B1Y & 0.1 &0.1 \\
	   & B1N & 0 &0.1 \\	   
    	   & B3Y & 0.3 &0.3 \\
	   & B3N & 0 &0.3 \\	 
    	   & B5Y & 0.5 &0.5 \\
	   & B5N & 0 &0.5 \\	 
    	 A2029........................& C0 &0&0\\
    	   & C1Y & 0.1 &0.1 \\
	   & C1N & 0 &0.1 \\	   
    	   & C3Y & 0.3 &0.3 \\
	   & C3N & 0 &0.3 \\	 
    	   & C5Y & 0.5 &0.5 \\
	   & C5N & 0 &0.5 \\	 
   	 \hline
   	  
   	\end{tabular}
   \end{minipage}
\end{table}

\subsection{Modeling of Thermal Conduction}
\label{sec:conduction}

$\boldsymbol{F}$ in equation (\ref{hydro3}) is the heat flux due to thermal conduction, and it may be written as
\begin{equation}
  \label{spitzercon}
  {\boldsymbol{F}} = -\kappa_{\text{eff}} \boldsymbol{\nabla}T \text{,}
\end{equation}
\noindent
where $\kappa_{\text{eff}}$ is the effective conductivity in the ICM. As in previous studies (e.g., \citealt{NarayanMedvedev2001}; \citealt{Komarov2016}), the suppression of conductivity by turbulent magnetic fields and kinetic-scale instabilities is often represented by the effective conductivity reduced from the Spitzer formula,
\begin{equation}
  \kappa_{\text{eff}} =   f \kappa_{\text{sp}}{,}
\end{equation}
\noindent
where $f$ ($0\leq f \leq 1$) is the conductivity suppression factor, and $\kappa_{\text{sp}}$ is the classic Spitzer Coulombic conductivity (\citealt{Spitzer1962}), 
\begin{equation}
 \label{kappa}
  \kappa_{\text{sp}} = \frac{1.84 \times 10^{-5}}{\text{ln} \lambda} T^{5/2}\text{erg~s}^{-1}\mathrm{K}^{-7/2}\text{cm}^{-1} {,}
\end{equation}
\noindent
where $\text{ln} \lambda \sim 40$ is the usual Coulomb logarithm for the ICM \citep{cowie77}.

Assuming spherical symmetry, our simulations focus on thermal conduction along the radial direction. Inner regions of galaxy clusters show positive temperature gradients, and are susceptible to the HBI, which tends to rearrange magnetic field lines to be perpendicular to the radial direction, suppressing radial conduction. In contrast, outer regions of galaxy clusters exhibit negative temperature gradients, and are susceptible to the MTI, which tends to rearrange magnetic field lines to be preferentially radial, enhancing radial conduction. Therefore we treat radial thermal conductivity in inner and outer regions of galaxy clusters separately:
\begin{eqnarray}
f= 
\begin{cases}
~f_{\rm in}      & \quad \text{when $dT/dr > 0$ (inner regions),}\\ 
~f_{\rm out}     & \quad \text{when $dT/dr \leq 0$ (outer regions).}
\end{cases}
\end{eqnarray}

To investigate the impact of thermal conduction on the ICM temperature profile, we performed a large suite of hydrodynamic simulations with various levels of thermal conductivity for three massive galaxy clusters: A1795, A478, and A2029. Table \ref{simulatable} lists our simulations with the adopted values of $f_{\rm in}$ and $f_{\rm out}$. As discussed in Section \ref{section1}, we focus on the outer regions of galaxy clusters with $dT/dr \leq 0$ (i.e., $r\geq r_{\rm peak}$, where $r_{\rm peak}$ is the radial location where the ICM temperature profile has the peak value; see Table \ref{table3}). The value of $f_{\rm out}$ is particularly important, and we consider three cases for $f_{\rm out}$:
\begin{itemize}
\item{$f_{\rm out}=0.5$ represents efficient radial conductivity in largely radial magnetic fields induced by the MTI, which is expected to develop if thermal conduction along field lines operates unimpeded \citep{Parrish2008}.}
\item{$f_{\rm out}=0.3$ is the intermediate case which may represent the level of thermal conductivity in a turbulent ICM \citep{NarayanMedvedev2001} induced by external processes (e.g., cosmic inflows).}
\item{$f_{\rm out}=0.1$ represents the maximum level of thermal conductivity if thermal conduction along magnetic field lines is suppressed by kinetic mirror or whistler instabilities. \citet{Komarov2016} demonstrate that mirror instability alone could suppress conductivity along field lines by a factor of $5$. Along the radial direction, further suppression of conduction may result from the existence of non-radial field lines. Considering the most favorable situation where field lines are preferentially radial due to the MTI, this additional suppression factor is about $0.5$, leading to $f_{\rm out}=0.1$. In reality when mirror or whistler instabilities operate, the MTI may not have time to develop and field lines are thus not preferentially radial, resulting in the value of $f_{\rm out}$ to be significantly less then $0.1$. In addition, \citet{clark16} argued that unstable whistler waves driven by the electron heat flux may suppress thermal conduction by an extreme factor of $10^{6}$ (but see \citealt{clark18} and \citealt{komarov17} for further discussions).}
\end{itemize}
\noindent
In addition, we also consider a case for each cluster where $f_{\rm out}=f_{\rm in}=0$ (e.g., run A0 for A1795), and the ICM in this case remains in its initial hydrostatic equilibrium state, as expected.

Inner cluster regions with $r\leq r_{\rm peak}$ contain cool cluster cores, where the temperature evolution is complicated by the effects of radiative cooling and active galactic nucleus (AGN) feedback (see, e.g., \citealt{mcnamara12}; \citealt{guo16};\citealt{guo18}), and is beyond the scope of the present work. Nonetheless, inward conduction in inner regions may affect the temperature evolution in outer regions, and therefore we consider two cases for the value of $f_{\rm in}$: $f_{\rm in}=f_{\rm out}$ or $0$. In the latter case, we manually turn off inward thermal conduction in inner regions with $dT/dr > 0$, and consider the impact of outward conduction alone on the outer temperature profile. Physically, the value of $f_{\rm in}$ is affected by the development of HBI and turbulence. For very weak magnetic fields, the HBI wraps field lines to be perpendicular to the radial direction, leading to $f_{\rm in} \sim 0$ \citep{parrish08b}. For intermediate magnetic fields, HBI-stable filaments form along the radial direction, enabling radial conduction at a level of $10\%$-$25\%$ of the Spitzer value (\citealt{Avara13}).

\subsection{Initial Conditions and Gravitational Potential}
\label{initcond}

  \begin{table*}
   \centering
   \begin{minipage}{120mm}
   	\caption{Parameters for Initial Gas Temperature Profiles in Our Galaxy Clusters}
   	\begin{tabular}{@{}l c c c c c c c c }
   	   \hline \hline
   	   &$T_0$ \footnote{The numerical values of the parameters listed in this Table are adopted from \citet{Vikhlinin2006}.} 
   	   &$r_t$
   	   &$a$
   	   &$b$
   	   &$c$
   	   &$T_{\rm min}/T_0$
   	   &$r_{\rm cool}$
   	   &$a_{\rm cool}$
   	   \\
	   Cluster
   	   &(keV)
   	   &(Mpc)
   	   &&&&&(kpc)
   	   &
   	   \\
   	   \hline
   	   A1795........................ & 9.68 & 0.55& 0.00 & 1.63& 0.9& 0.10& 77& 1.03
   	 \\A478......................... & 11.06 &0.27 &0.02 &5.00 &0.4 &0.38 &129 &1.60
   	 \\A2029........................ & 16.19 & 3.04 &-0.03 &1.57 &5.9 &0.10 &93 &0.48	 
   	 \\
   	 \hline
   	       	\label{tempara} 
     \end{tabular}
   \end{minipage}
 \end{table*}

  \begin{table}
   \centering
   \begin{minipage}{90mm}
   	\caption{Various Properties of Our Galaxy Clusters}
   	\label{table3}
   	\begin{tabular}{@{}l c c c c c c}
   	 \hline\hline
   	 &$r_{500}$ \footnote{The values of $r_{500}$ and $M_{500}$ are adopted from \citet{Vikhlinin2006}. \label{footnotea}} 
   	 &$M_{500}$\textsuperscript{\ref{footnotea}}
   	 &$r_{\rm peak}$ \footnote{$r_{\rm peak}$ is the radial location where the initial temperature profile has the peak value.} 
   	 &$T_{\rm peak}$\footnote{$T_{\rm peak}$ is the maximum temperature in the initial ICM temperature profile.} 
   	 &$0.3r_{500}$	 
   	 &$T_{0.3r500}$ \footnote{$T_{0.3r500}$ is the initial gas temperature at $r=0.3r500$.}
   	 \\
   	 Cluster	 
   	 &(kpc)
   	 &$(10^{14}M_{\odot})$
   	 &(kpc)
   	 &(keV)
   	 &(kpc)	 
   	 &(keV)	 
   	 \\
   	 \hline
   	 A1795 & 1235 &6.03  &258 &6.73 &370.5 & 6.54
   \\A478  & 1337 &7.68  &266 &8.95 &401.1 &8.62
    \\A2029 & 1362 &8.01 &329 &9.24 &408.6 & 9.22  
   \\
   \hline
   	\end{tabular}
   \end{minipage}
  \end{table}

\begin{table*}
  \centering
  \begin{minipage}{120mm}
  	\caption{Parameters for Initial Gas Density Profiles in Our Galaxy Clusters}
  	\begin{tabular}{@{} l c c c c c c c c c c}
  	 \hline
	 \hline
  	     &$n_{0}$ \footnote{The numerical values of the parameters listed in this Table are adopted from \citet{Vikhlinin2006}.} 
  	     &$r_c$ 
  	     &$r_s$ 
  	     &$\alpha$
  	     &$\beta$
  	     &$\varepsilon$
  	     &$n_{02}$
  	     &$r_{c2}$ 
  	     &$\beta_2$
  	     \\
	     Cluster
  	     &$10^{-3}$~cm$^{-3}$
  	     &(kpc)
  	     &(kpc)
  	     &&&&$10^{-1}$~cm$^{-3}$
  	     \\
  	  \hline
  	  A1795 & 31.175 &38.2 & 682.5 & 0.195 & 0.491 & 2.606 & 5.695 & 3.00
  	  & 1.000\\
  	  A478 & 10.170 & 155.5 & 2928.9 & 1.254 & 0.704 & 5.000 & 0.762 & 
  	  23.84 & 1.000\\
  	  A2029 & 15.721 & 84.2 & 908.9 & 1.164 & 0.545 & 1.669 & 3.510 & 
  	  5.00 & 1.000\\	  
  	\hline 
	   \label{denpara}	     
  	\end{tabular}
  \end{minipage}
\end{table*}

The initial conditions of our clusters are directly adopted from {\it Chandra} X-ray observations presented in \citet{Vikhlinin2005} and \citet{Vikhlinin2006}. For the initial temperature profile, we adopt the analytic model in \citet{Vikhlinin2006}, which fits very well the deprojected 3-dimensional temperature distribution measured by {\it Chandra}:
  \begin{equation}
    T(r) = T_0\frac{(x+T_{\rm min}/T_0)}{(x+1)} \frac{(r/r_{t})^{-a}}{[1+(r/r_t)^b]^{c/b}}~{,}   \label{temequ}
 \end{equation}
\noindent
where 
  \begin{equation}
x=(\frac{r}{r_{\rm cool}})^{a_{\rm cool}}~{.} 
 \end{equation}
 \noindent
The values of the parameters in the above temperature model for each cluster are directly adopted from \citet{Vikhlinin2006}, and listed in Table \ref{tempara}. The resulting initial temperature profiles of our clusters are shown in the left panel of Figure \ref{plot1}. In the right four columns of Table \ref{table3}, we list the values of $r_{\rm peak}$, $T_{\rm peak}$, $0.3r500$, and $T_{0.3r500}$ for each cluster. Here $T_{\rm peak}=T(r_{\rm peak})$ and $T_{0.3r500}=T(0.3r500)$ are, respectively, the peak temperature and the temperature at $r=0.3r500$ in the initial ICM temperature profile.

For initial gas density profiles, we adopt the analytic expression for the observed emission measure profile in \citet{Vikhlinin2006}, which provides an adequately good fit to {\it Chandra} data for all our clusters:
\begin{equation}
  \label{denequ}
  \begin{split}
n_{\rm e}n_{\rm p} = n_{0}^{2}\frac{(r/r_c)^{-\alpha}}{(1+r^2/r_c^2)^{3\beta-\alpha/2}}
           & \frac{1}{(1+r^3/r_s^3)^{\varepsilon/3}}+\\
           \frac{n_{02}^2}{(1+r^2/r_{c2}^{2})^{3\beta_2}} ~{,} 
   \end{split}
\end{equation}
where $n_{\rm p}$ is the proton number density. The initial gas density profile could be derived via $\rho=1.27m_{\rm p}(n_{\rm e}n_{\rm p})^{1/2}$, where $m_{\rm p}$ is the proton mass. The numerical values of density parameters in equation (\ref{denequ}) for our clusters are directly adopted from \citet{Vikhlinin2006}, and listed in Table \ref{denpara}. The resulting initial electron number density profiles of our clusters are shown in the right panel of Figure \ref{plot1}.

We assume that the ICM is initially in hydrostatic equilibrium, i.e., $\rho \nabla \Phi = -\nabla P$, which is used to solve for the gravitational potential $\Phi$. \citet{Vikhlinin2006} show that the resulting $\Phi$ profile is consistent with that contributed by a total density distribution of the Navarro-Frenk-White (NFW) profile (\citealt{Navarro1996}; \citealt{Navarro1997}). The second and third columns of Table \ref{table3} list the numerical values of $r_{500}$ and $M_{500}$, respectively, of the corresponding NFW profile for each of our clusters. Except for central regions dominated by brightest cluster galaxies, $\Phi$ in galaxy clusters is dominantly contributed by the dark matter distribution, and for simplicity, we assume that $\Phi$ is fixed during our simulations.

 \begin{figure}
  \includegraphics[width = 0.45\textwidth]{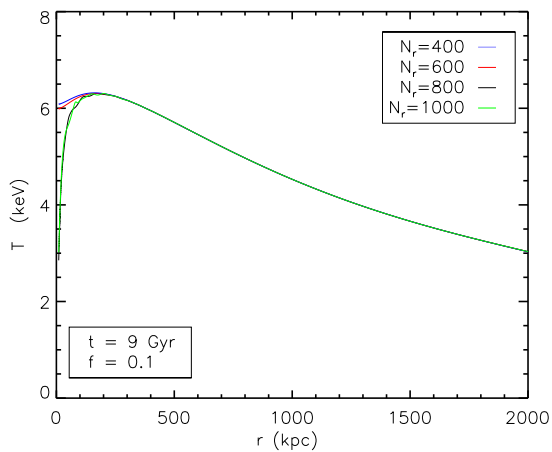}
   \caption{The radial temperature profile of the cluster A1795 at $t=9$ Gyr in a series of $4$ simulations with different radial resolutions $N_{r}=400$ (purple), $600$ (red), $800$ (black), and $1000$ (green). The conduction suppression factor in these runs is assumed to be $f_{\rm in}=f_{\rm out}=0.1$. The outer temperature distribution beyond its peak value converges in all these runs, while the inner temperature distribution converges in runs with $N_{r} = 800$ and $1000$.}
    \label{resolutionfig}
\end{figure}

\subsection{Simulation Setup}
 \label{sec:setup}
 
Assuming spherical symmetry, we solve equations (\ref{hydro1}) - (\ref{hydro3}) for the ICM by using the one-dimensional mode of the ZEUS-3D hydrodynamic code (\citealt{stone92}; \citealt{Clarke1996}; \citealt{Clarke2010}). This particular version of the code has been successfully used in several previous studies (e.g., \citealt{guo08}; \citealt{guo09}; \citealt{guo14a}; \citealt{guo14b}). 

Our computational domain extends from an inner boundary at $r_{\rm min} = 10$ kpc to an outer boundary at $r_{\rm max} = 2$ Mpc. We have also tried with smaller inner boundaries and larger outer boundaries, and found that our conclusions do not change. We adopt a logarithmically spaced grid with $(\Delta r)_{i+1}/(\Delta r)_{i} = (r_{\rm max}/r_{\rm min})^{1/N_{r}}$, where $N_{r} = 800$ is the total number of active zones. We performed a convergence test for the radial resolution in the cluster A1795 with $f_{\rm in}=f_{\rm out}=0.1$, and the results are shown in Figure \ref{resolutionfig}. The outer temperature profile beyond its peak value converges very well in all our test runs with $N_{r} \geq 400$, while the inner temperature profile converges in runs with $N_{r} = 800$ and $1000$. In all the simulations presented in the rest of the paper, the default resolution is chosen to be $N_{r} = 800$, corresponding to a smallest grid size $\Delta r=66$ pc at the cluster center. At the inner radial boundary, we choose the regular outflow boundary condition \citep{stone92}. At the outer boundary $r_{\rm max} = 2$ Mpc (close to the virial radius of our clusters), we assume that the gas density and temperature are fixed with time, as they may be maintained by cosmic accretion of baryons into the cluster potential (also see \citealt{guo08} and \citealt{guo09}). We start our simulations at $t=0$ and typically stop at $t=7$ Gyr.

\section{RESULTS}
\label{section:results}

We performed a large suite of hydrodynamic simulations for three representative clusters: A1795, A478, and A2029, and here we present our main results in this section. In Section \ref{sec:Tevol}, we investigate the evolution of the ICM temperature profile in our simulations with different levels of conductivity. We discuss the relevant implications on the level of thermal conductivity and the potential importance of kinetic-scale instabilities in real clusters in Section \ref{sec:implications}.

\subsection{Evolution of the ICM Temperature Profile}
\label{sec:Tevol}

\begin{figure*}
	\centering
	\begin{center}
		\includegraphics[width=0.3\textwidth]{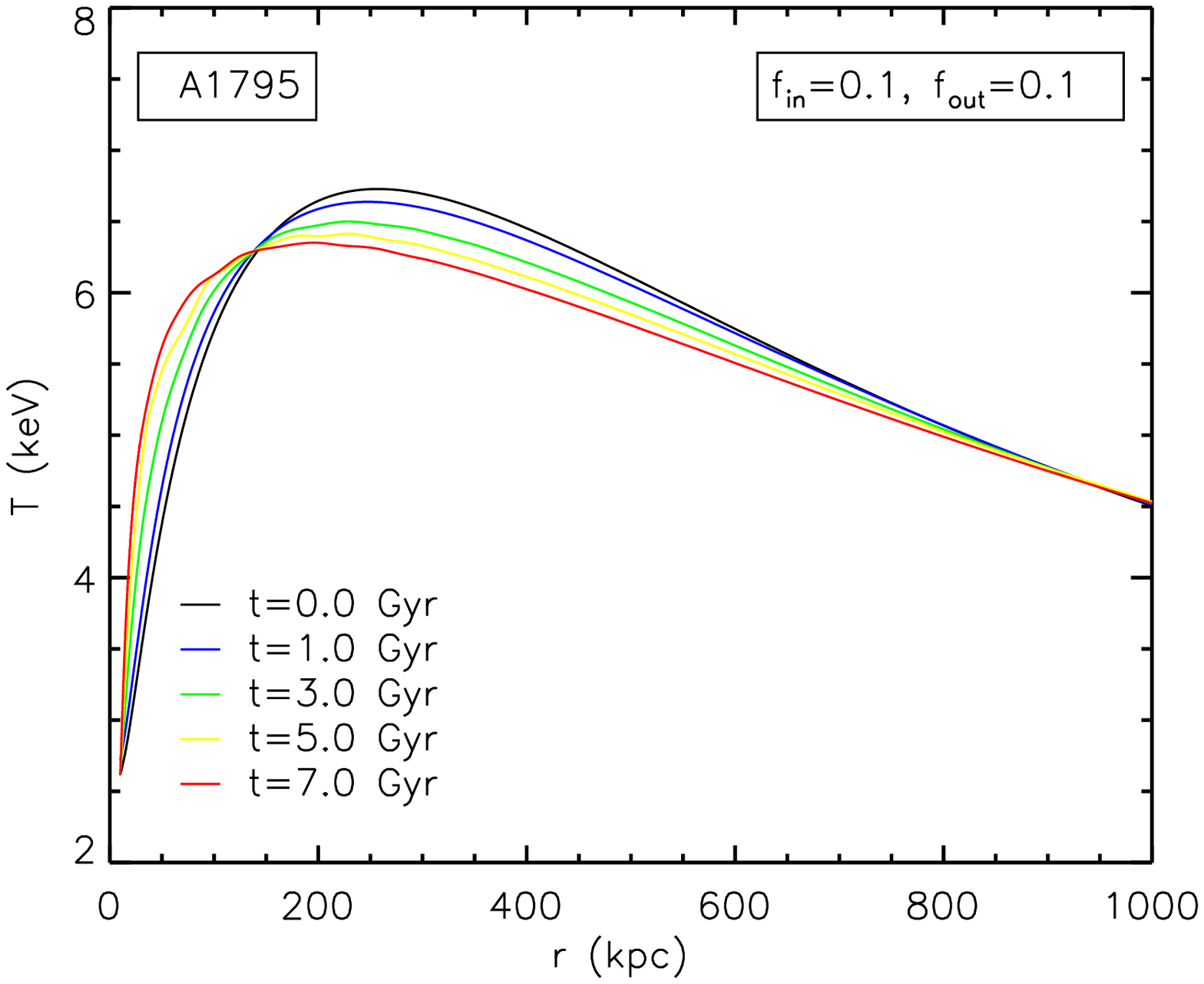}\hspace{0.27cm}
		\includegraphics[width=0.3\textwidth]{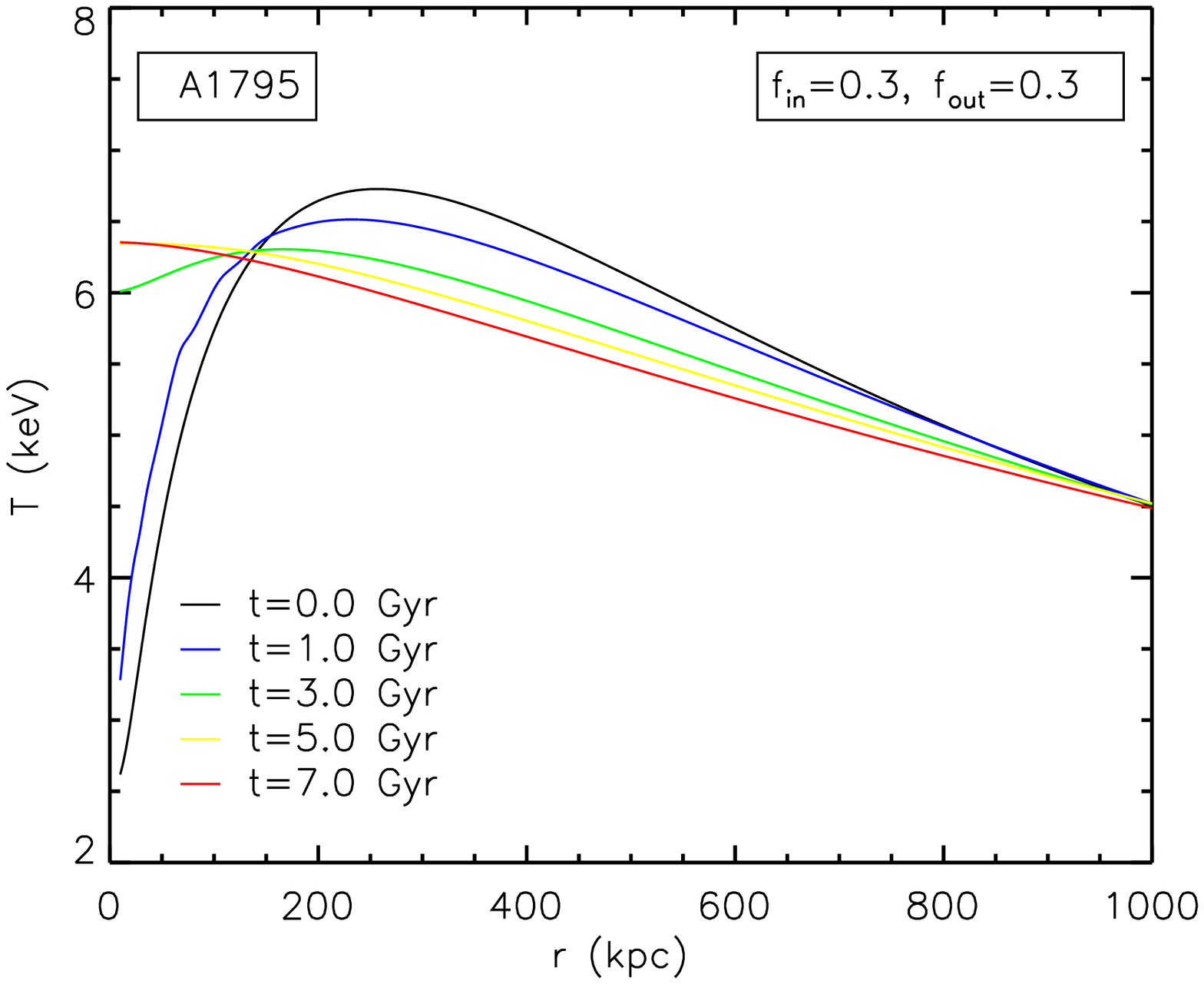}\hspace{0.27cm}
		\includegraphics[width=0.3\textwidth]{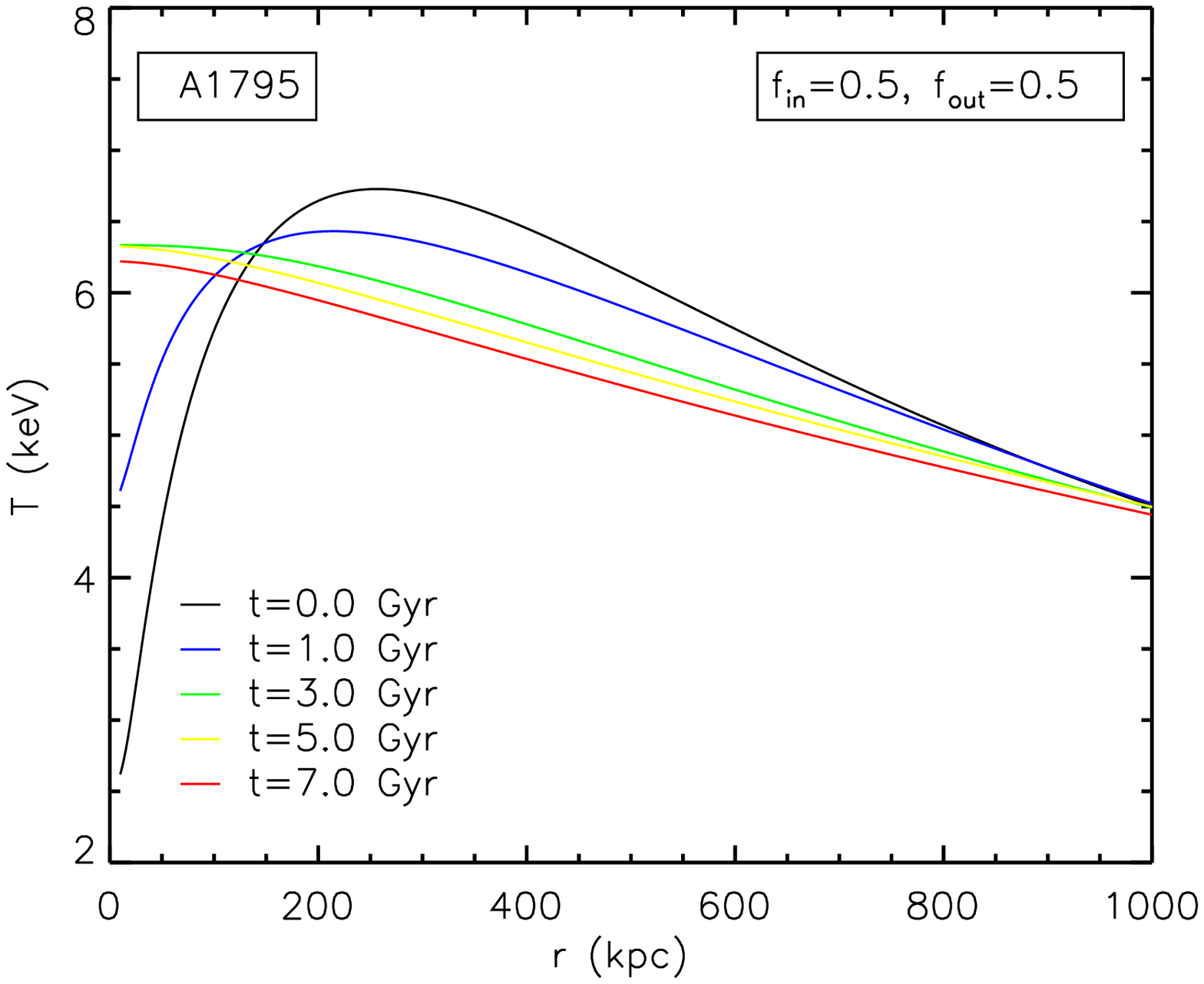}\hspace{0.27cm}\\
		\vspace{-0.5cm}
	\end{center}
	\begin{center}
		\includegraphics[width=0.3\textwidth]{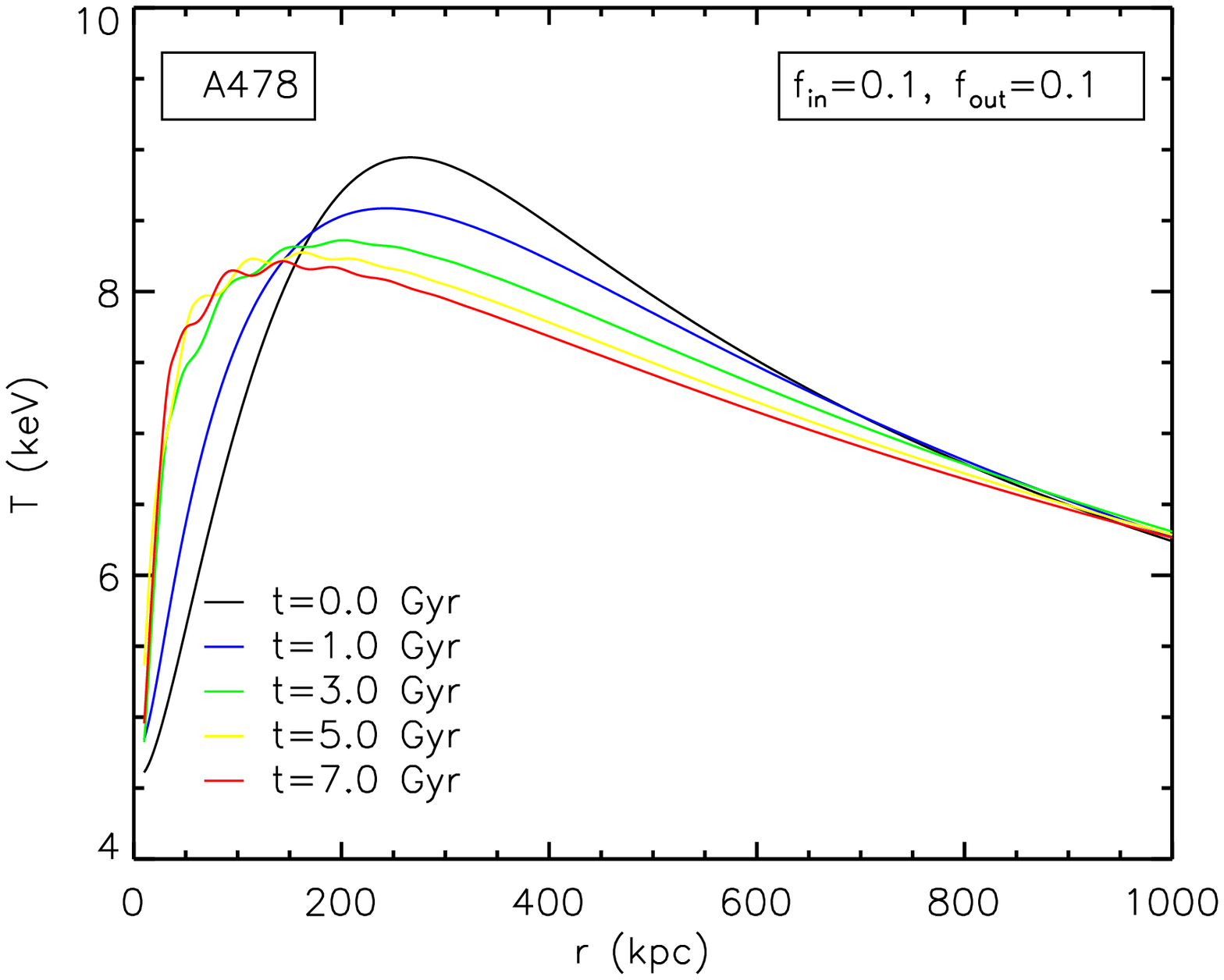}\hspace{0.27cm}
		\includegraphics[width=0.3\textwidth]{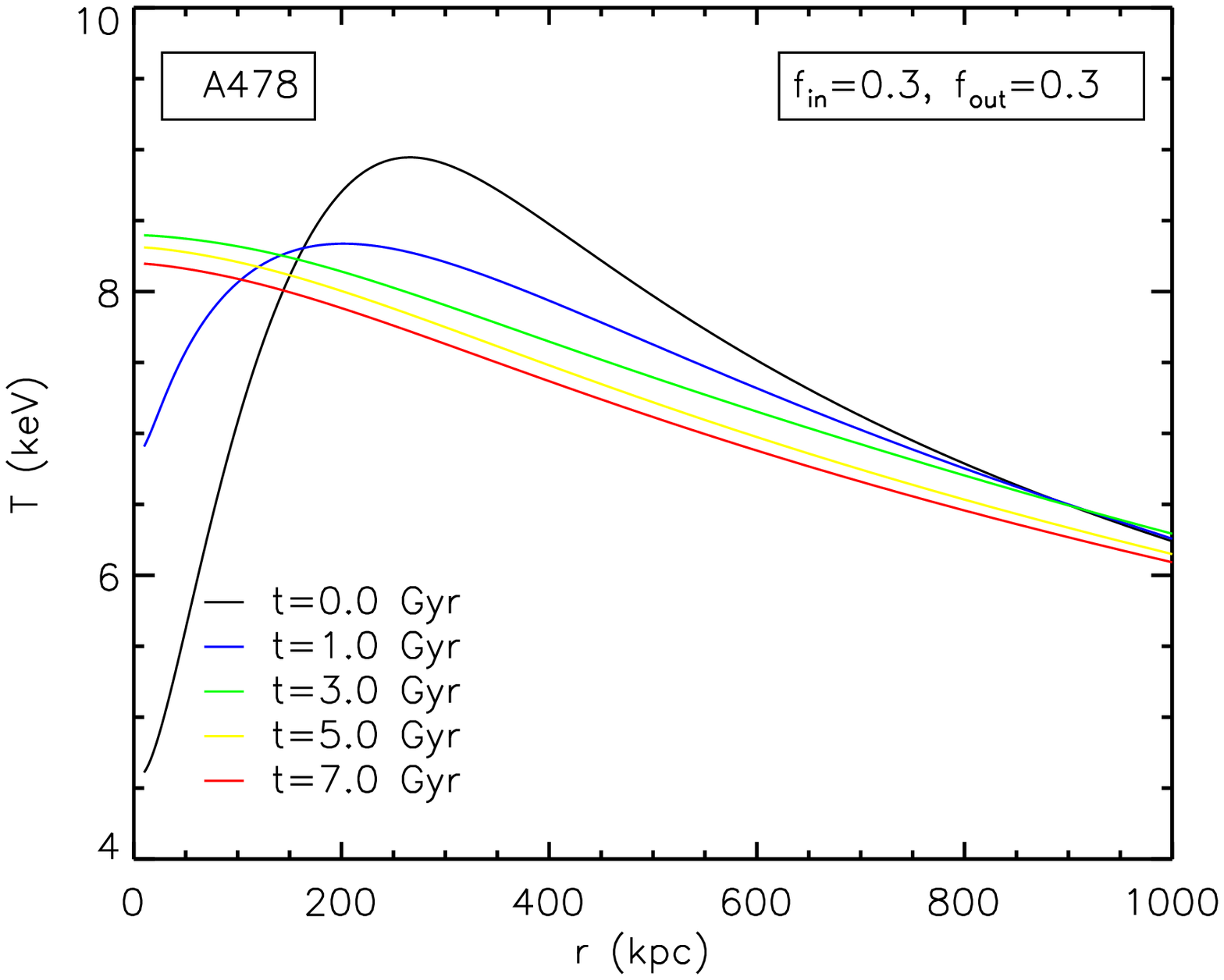}\hspace{0.27cm}
		\includegraphics[width=0.3\textwidth]{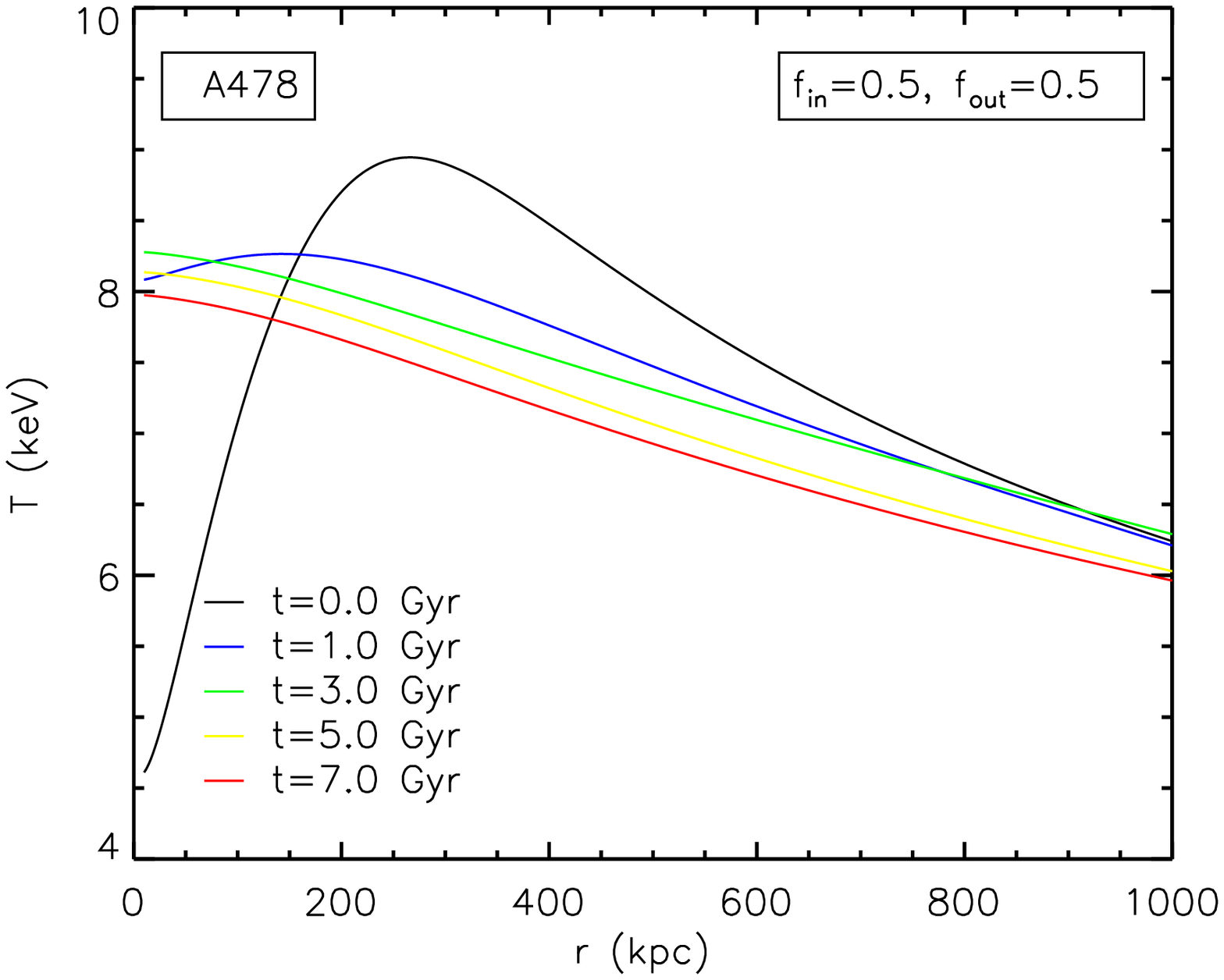}\hspace{0.27cm}\\
		\vspace{-0.5cm}		
	\end{center}
	\begin{center}
		\includegraphics[width=0.3\textwidth]{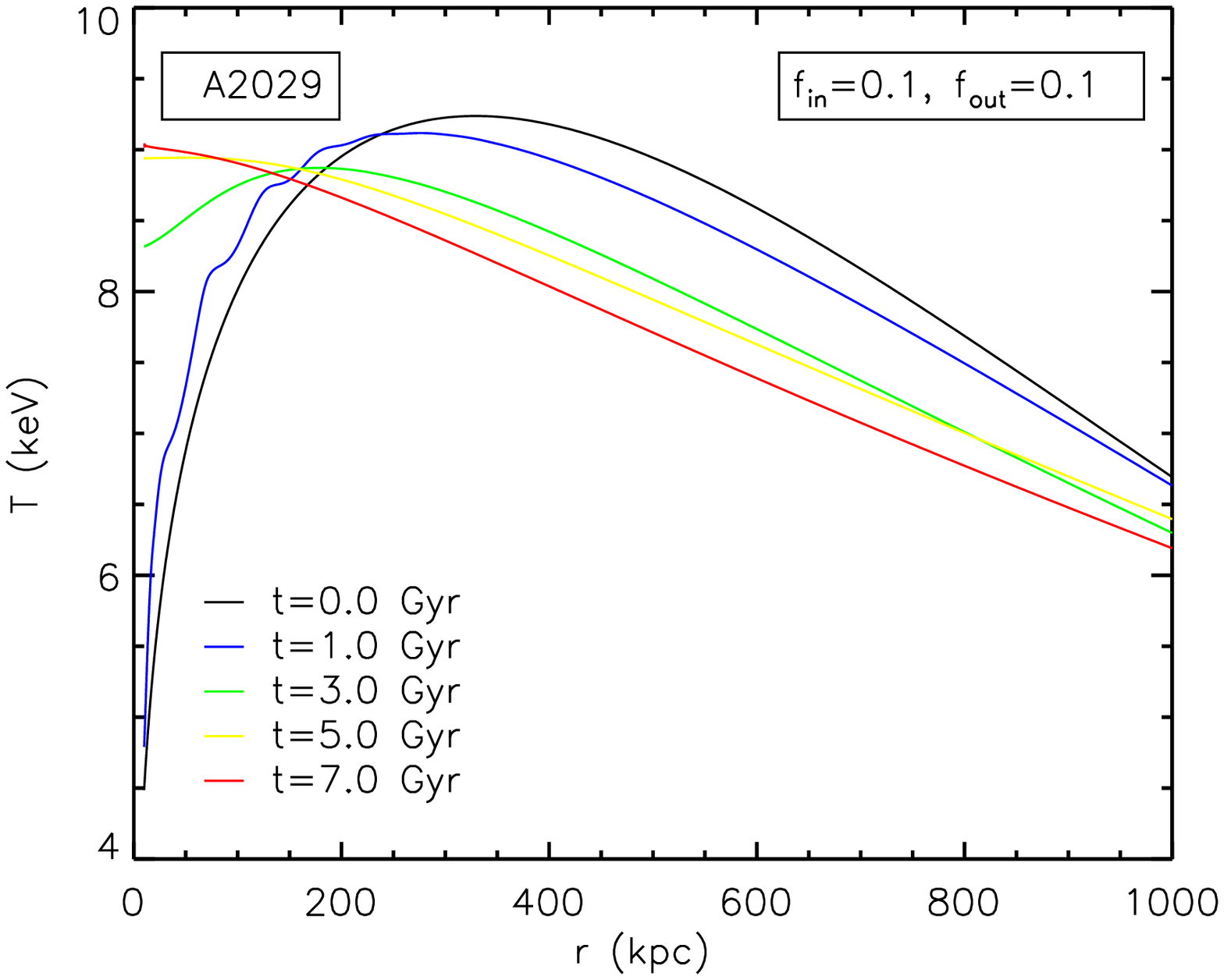}\hspace{0.27cm}
		\includegraphics[width=0.3\textwidth]{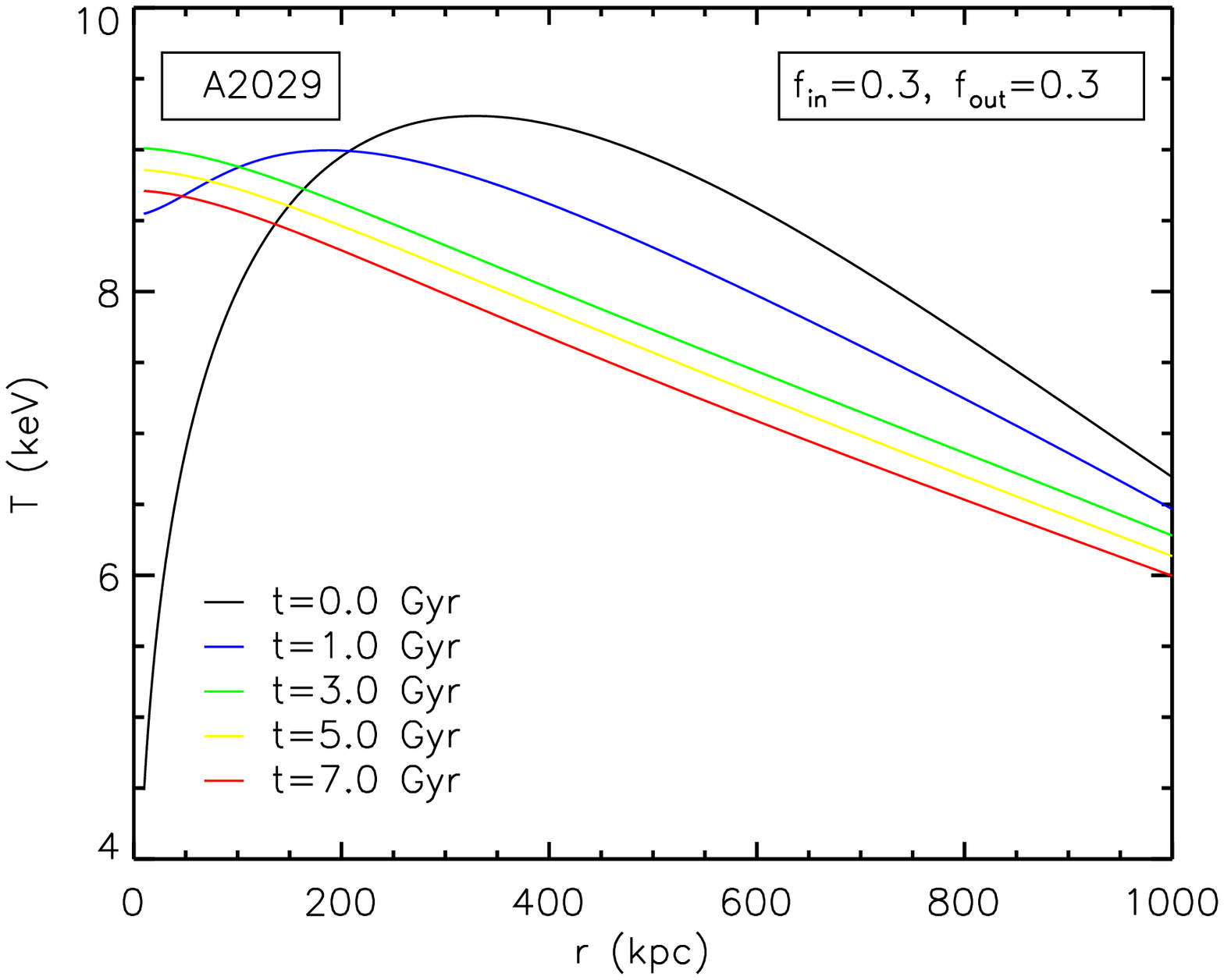}\hspace{0.27cm}
		\includegraphics[width=0.3\textwidth]{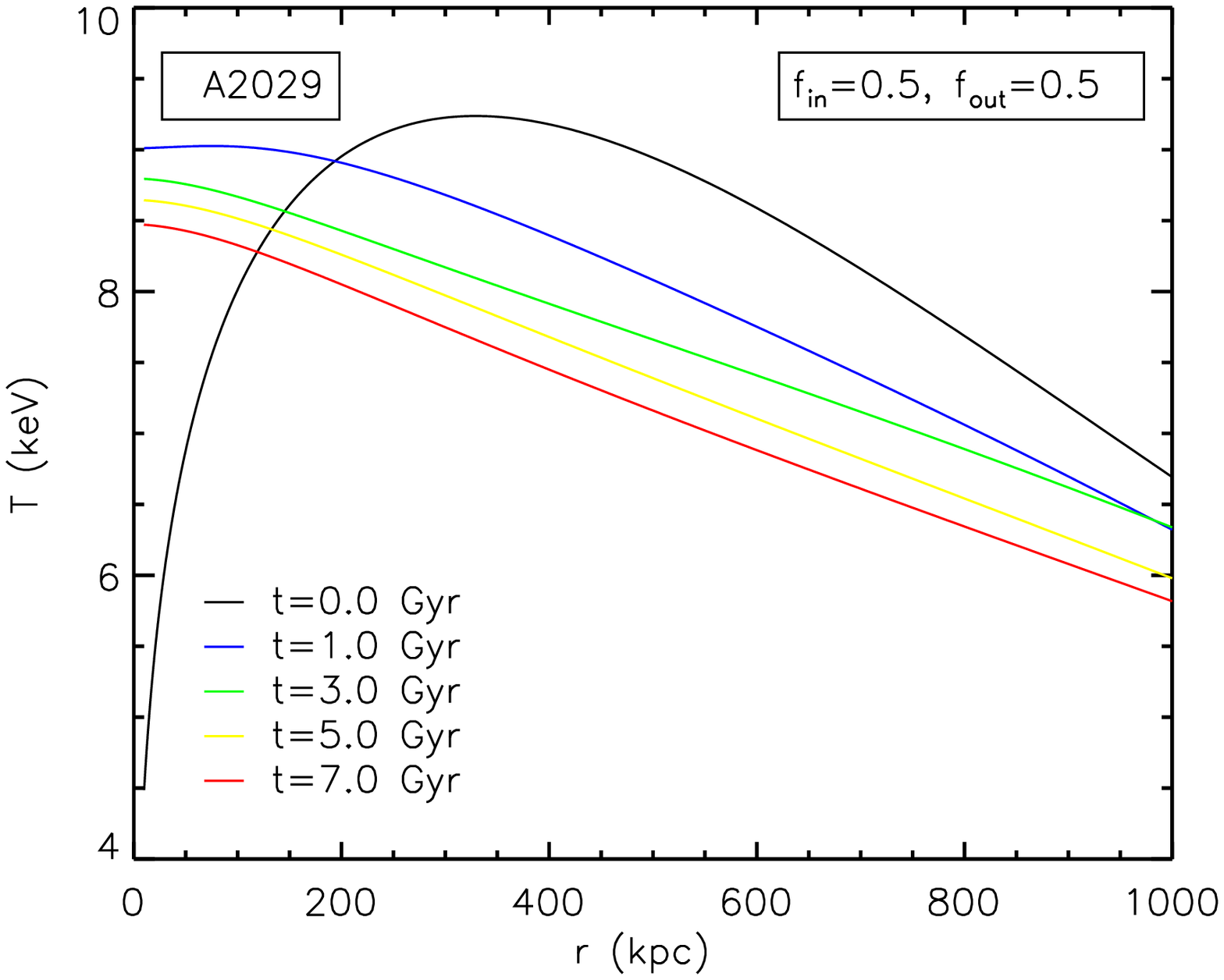}\hspace{0.27cm}\\
	\end{center}	
        \vspace{-0.4cm}
	\caption{Evolution of the ICM temperature profile in the cluster A1795 (top row), A478 (middle row), and A2029 (bottom row) with the conductivity suppression factor $f=0.1$ (left column), $0.3$ (middle column), and $0.5$ (right column). In these simulations, $f_{\rm in}=f_{\rm out}$.}
	\label{tempevo}
\end{figure*}

We first look at runs A0, B0, and C0 (for the cluster A1795, A478 and A2029, respectively), where thermal conduction is turned off (i.e., $f_{\rm in}=f_{\rm out}=0$). In each of these runs, the ICM temperature and density profiles do not evolve with time during the whole simulation, confirming that the clusters in our simulations are indeed initially in hydrostatic equilibrium.

\begin{figure*}
  \centering
  \begin{center}
   \includegraphics[width=0.3\textwidth]{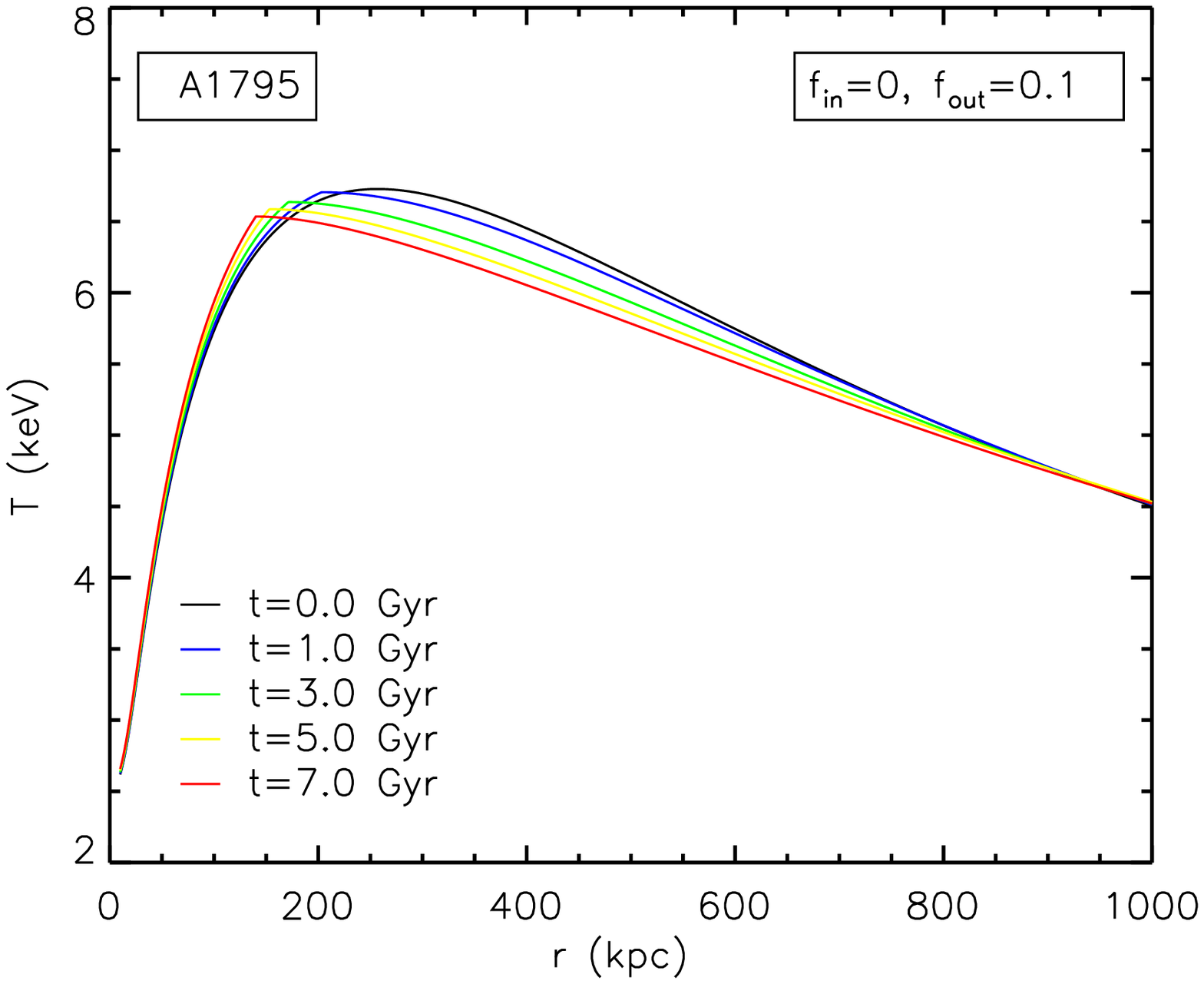}\hspace{0.25cm}
   \includegraphics[width=0.3\textwidth]{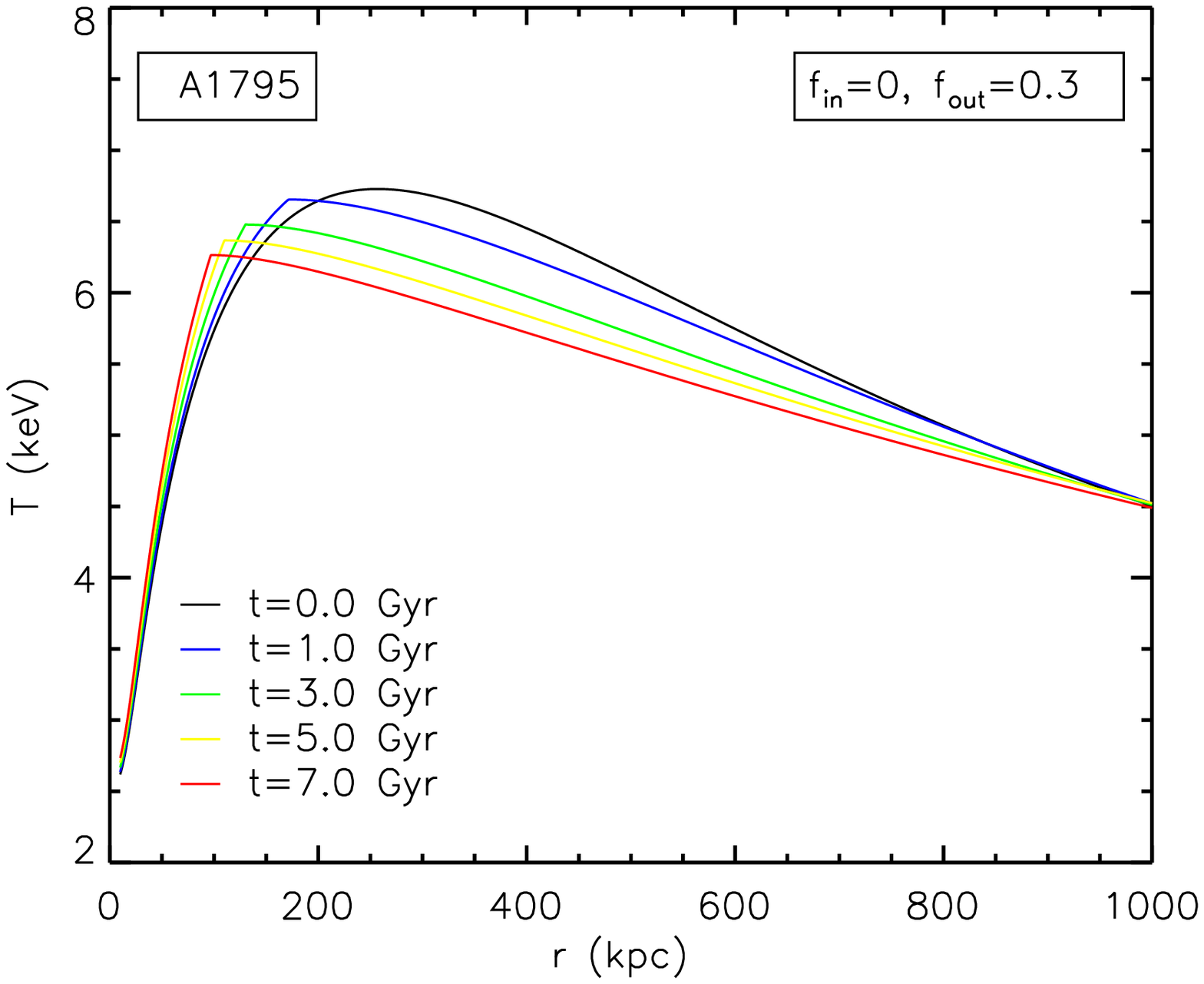}\hspace{0.25cm}
   \includegraphics[width=0.3\textwidth]{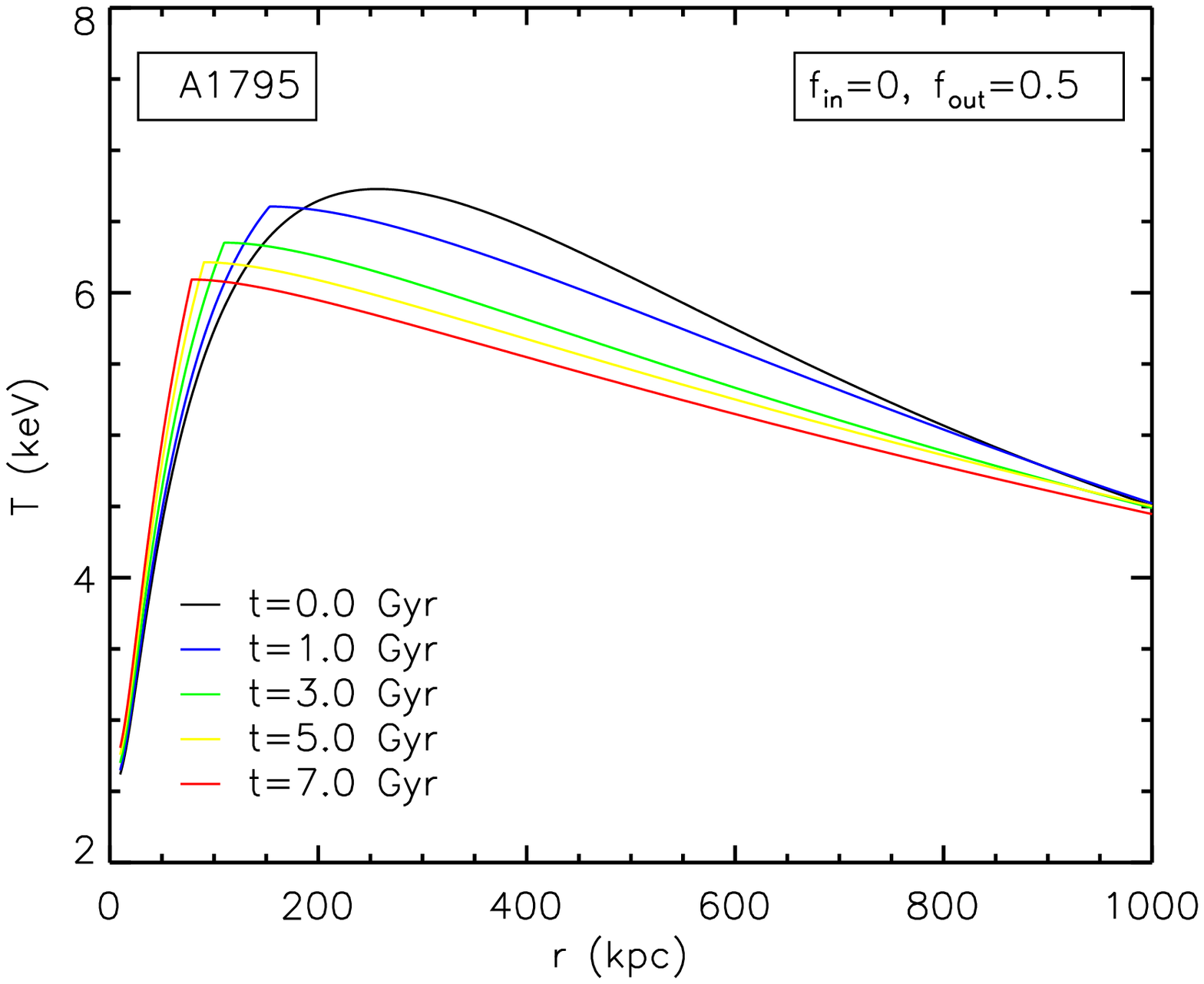}\hspace{0.25cm}\\
   		\vspace{-0.5cm}
  \end{center}
\begin{center}
	\includegraphics[width=0.3\textwidth]{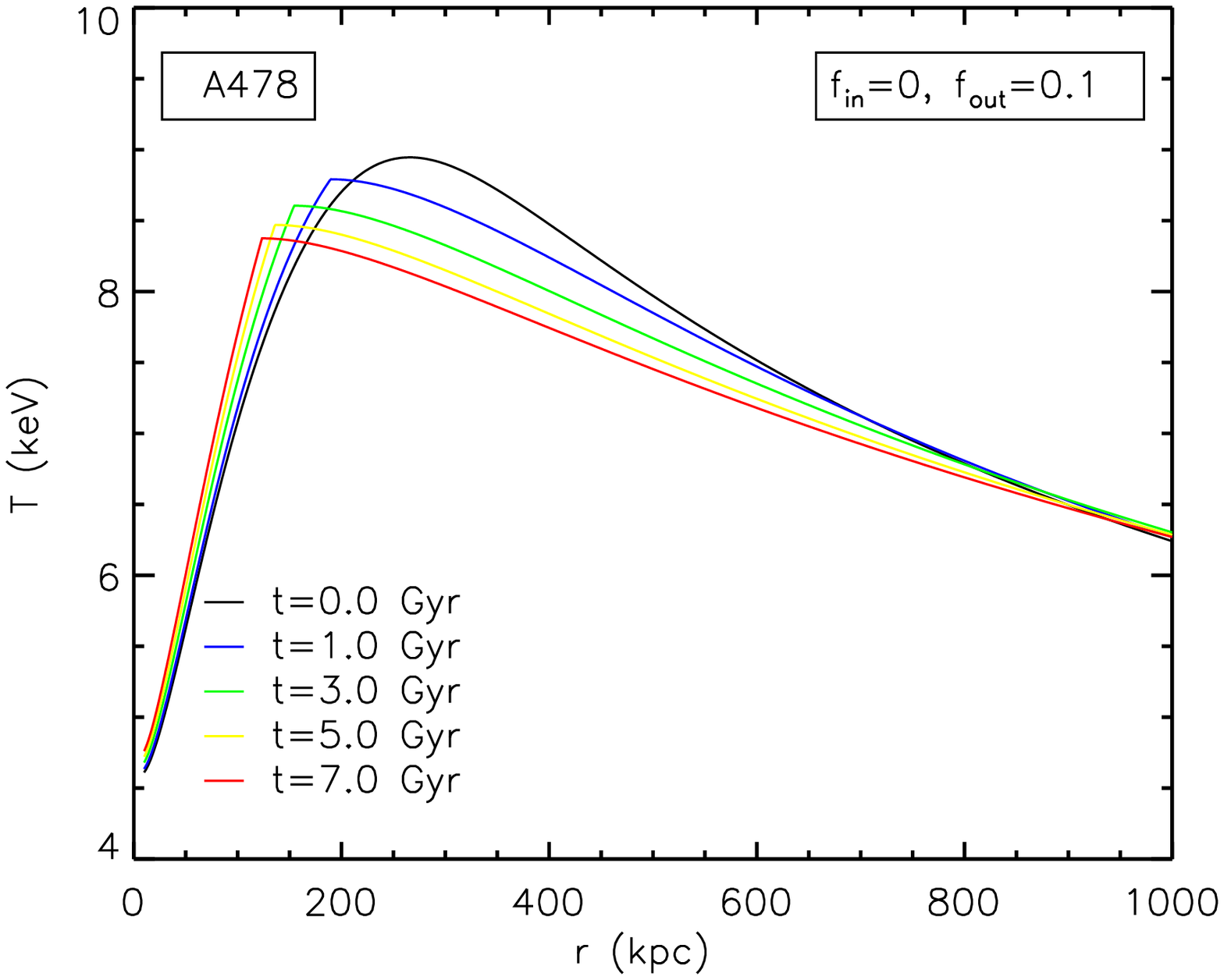}\hspace{0.25cm}
	\includegraphics[width=0.3\textwidth]{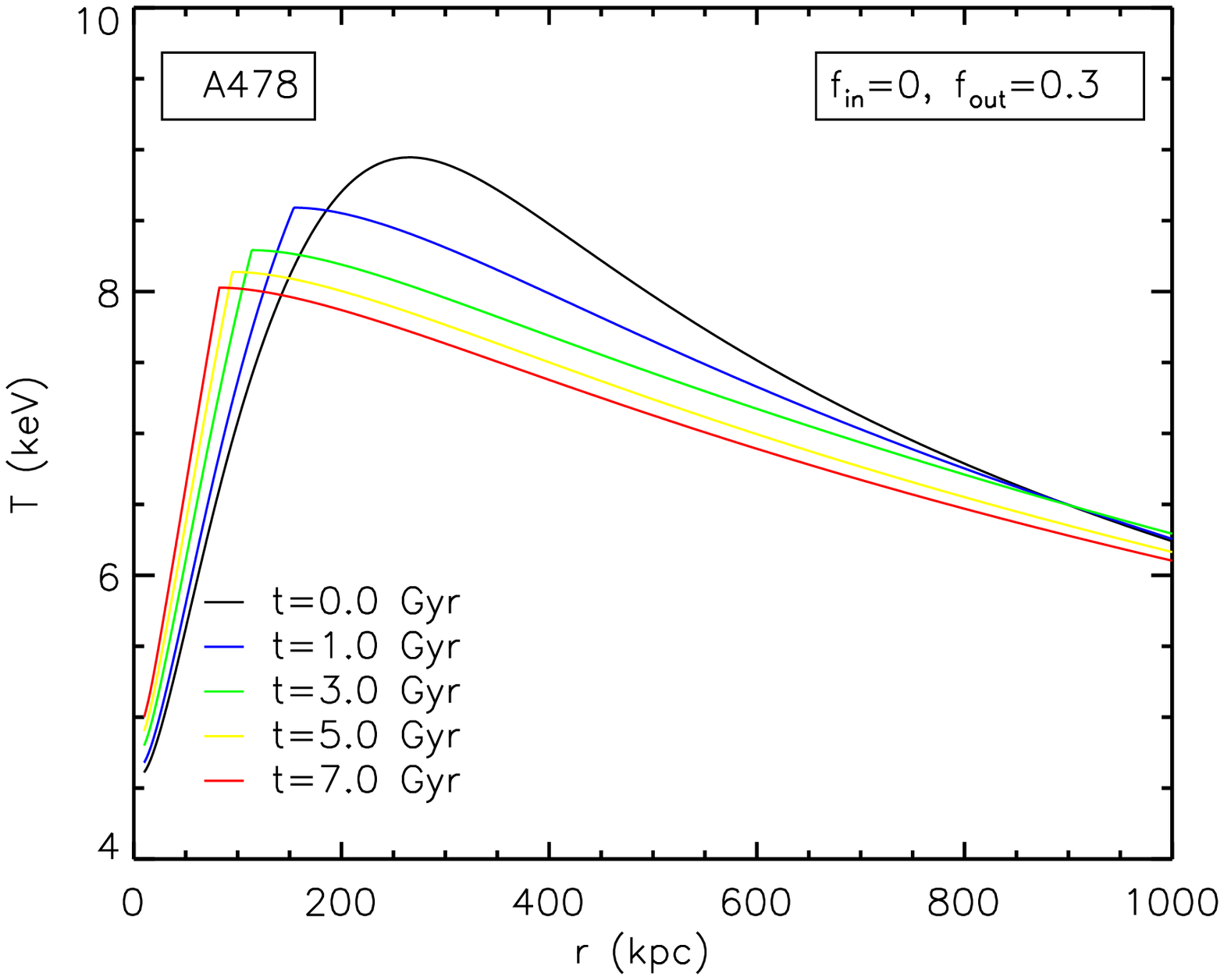}\hspace{0.25cm}
	\includegraphics[width=0.3\textwidth]{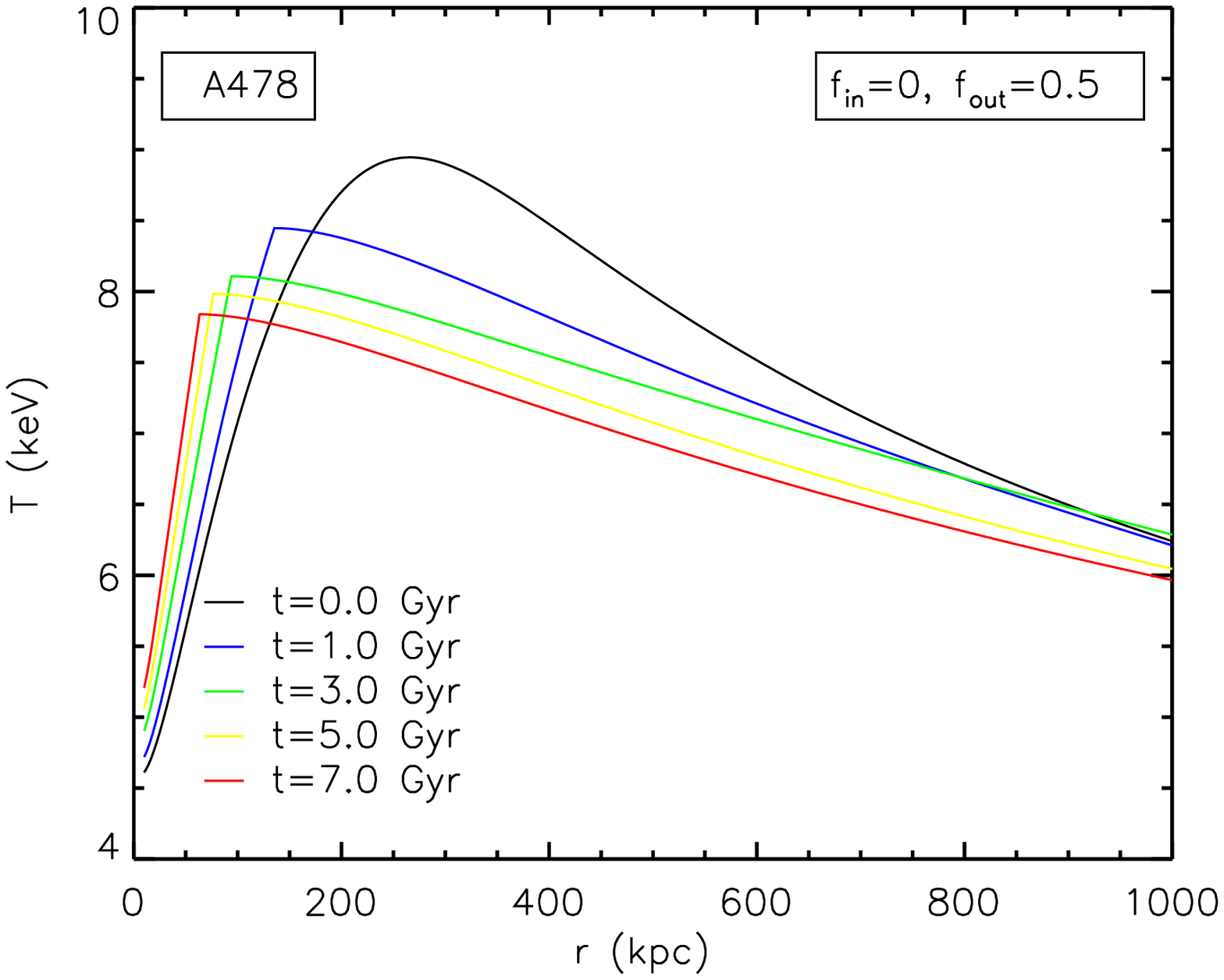}\hspace{0.25cm}\\
			\vspace{-0.5cm}
\end{center} 
 \begin{center}
	\includegraphics[width=0.3\textwidth]{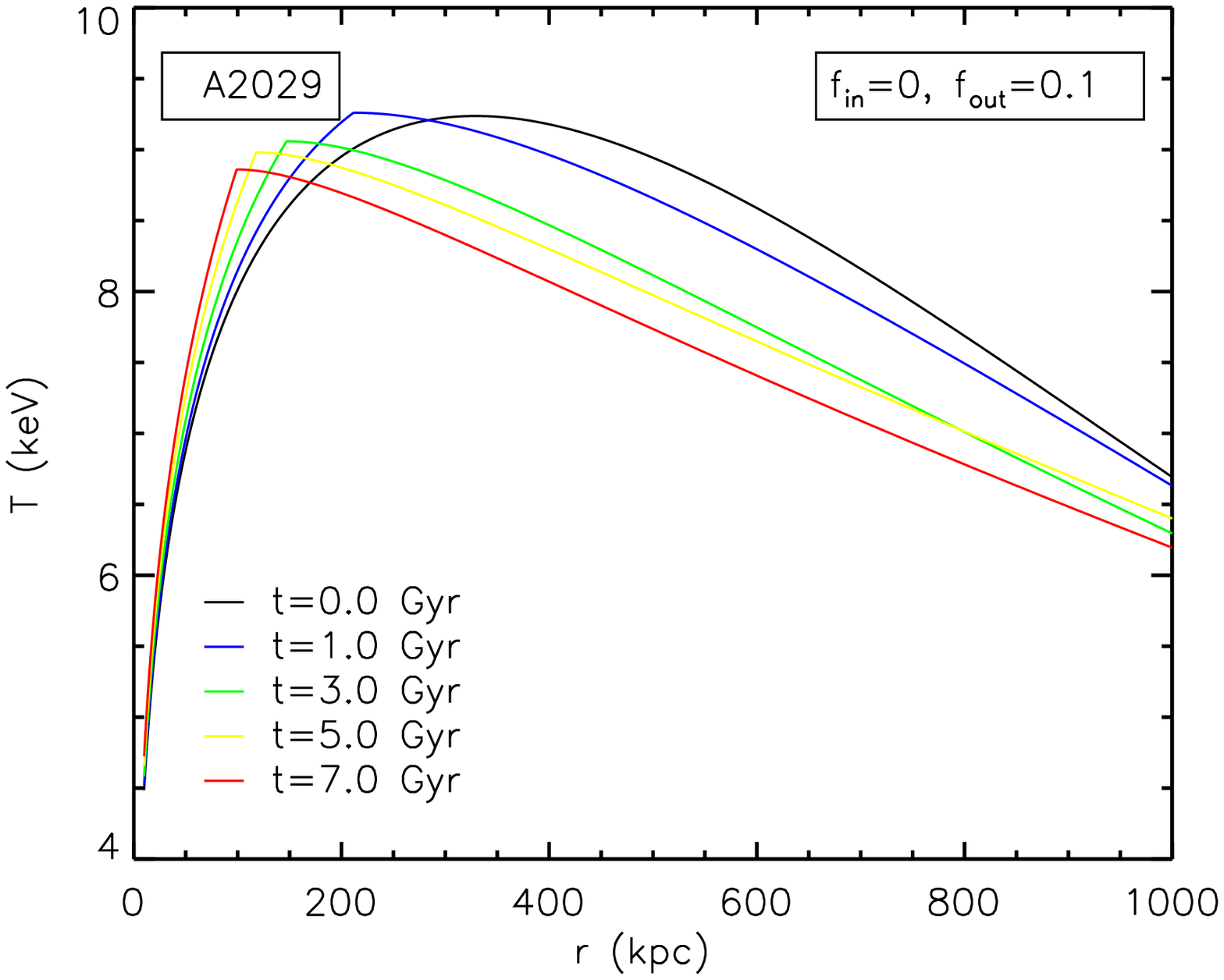}\hspace{0.25cm}
	\includegraphics[width=0.3\textwidth]{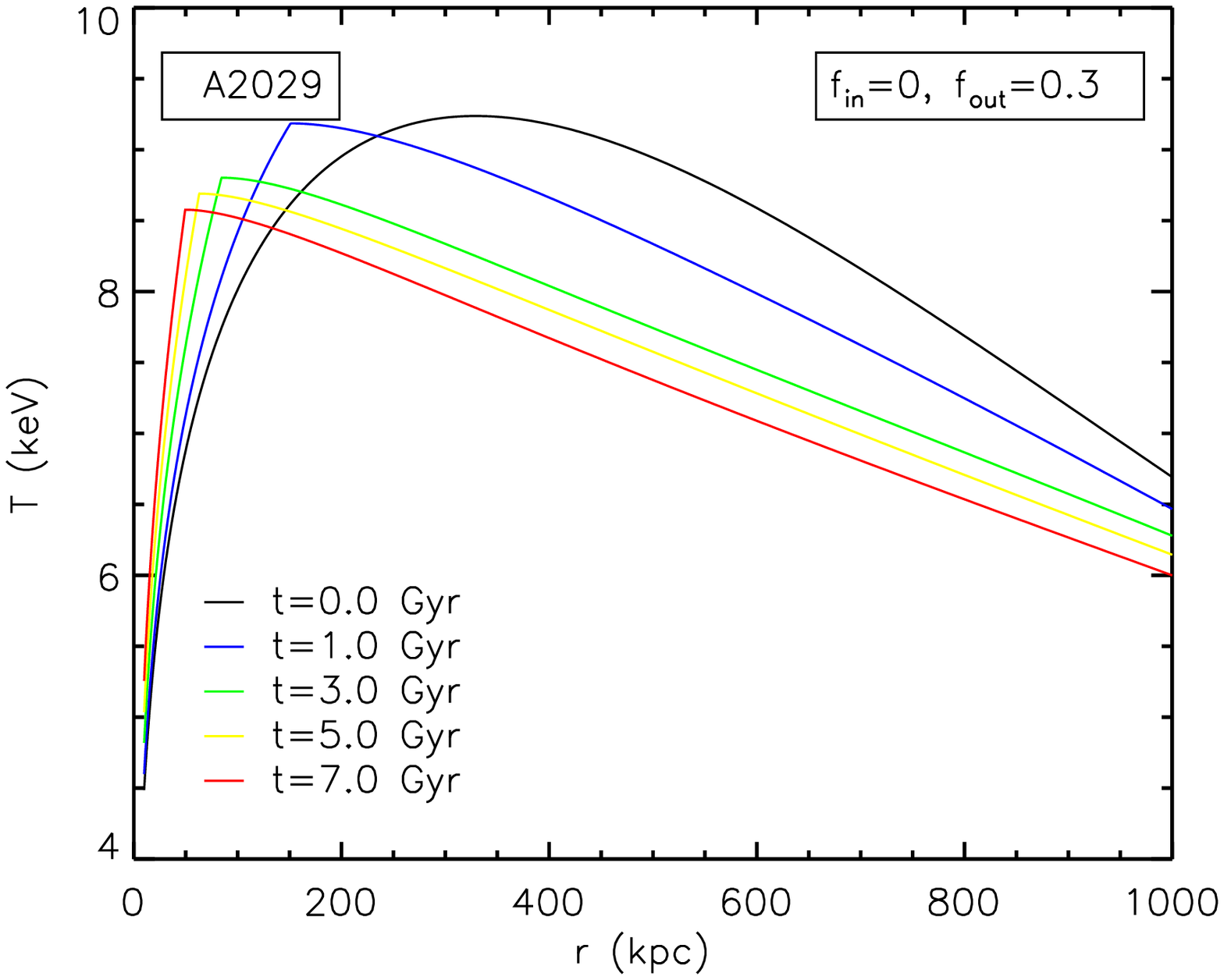}\hspace{0.25cm}
	\includegraphics[width=0.3\textwidth]{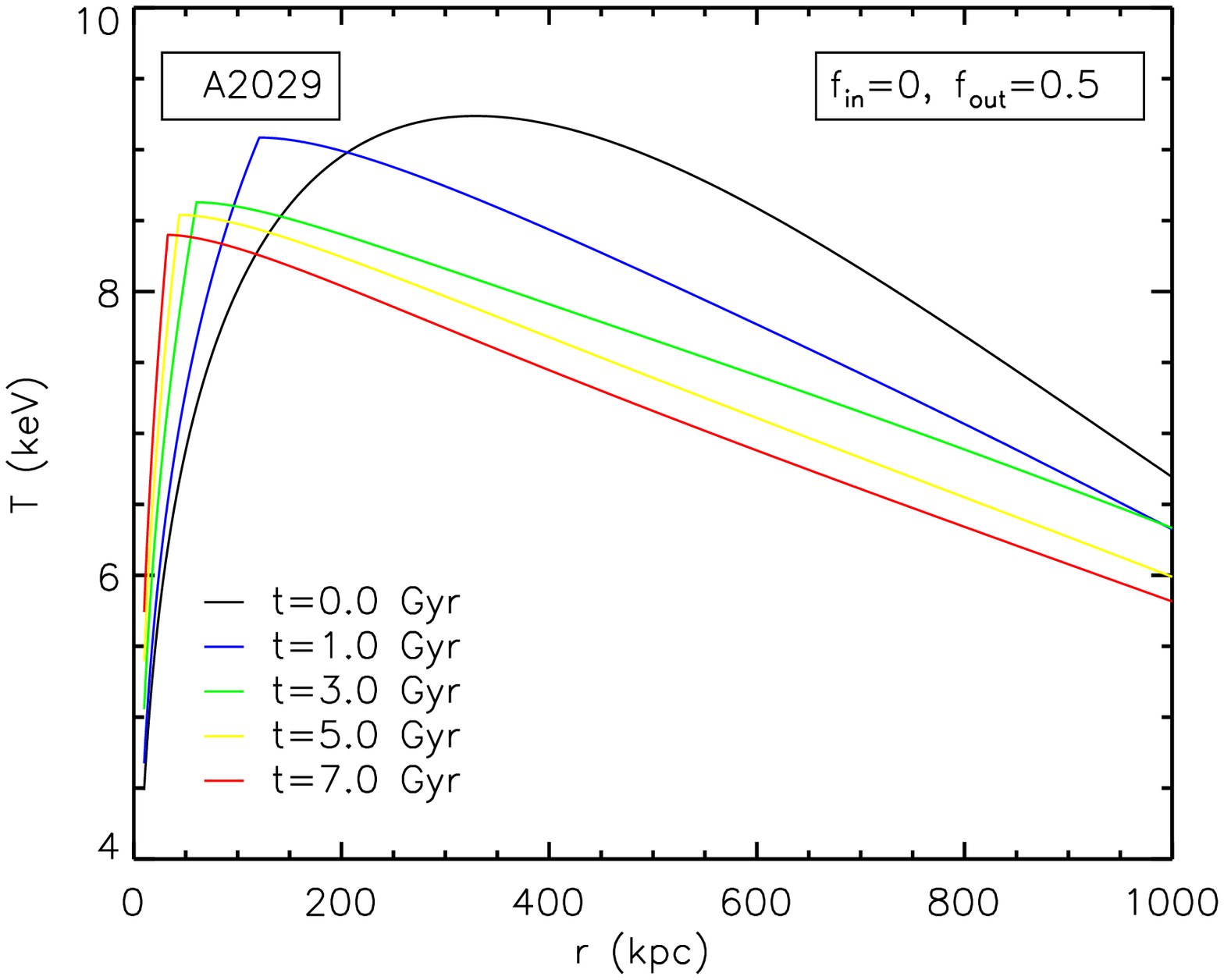}\hspace{0.25cm}
\end{center}
        \vspace{-0.4cm}
  \caption{Same as Fig. \ref{tempevo}, except for runs where inward thermal conduction in inner regions with $dT/dr>0$ is prohibited (i.e. $f_{\rm in}=0$).}
  \label{tempnic}
\end{figure*}

\begin{figure}
  \includegraphics[width=0.45\textwidth]{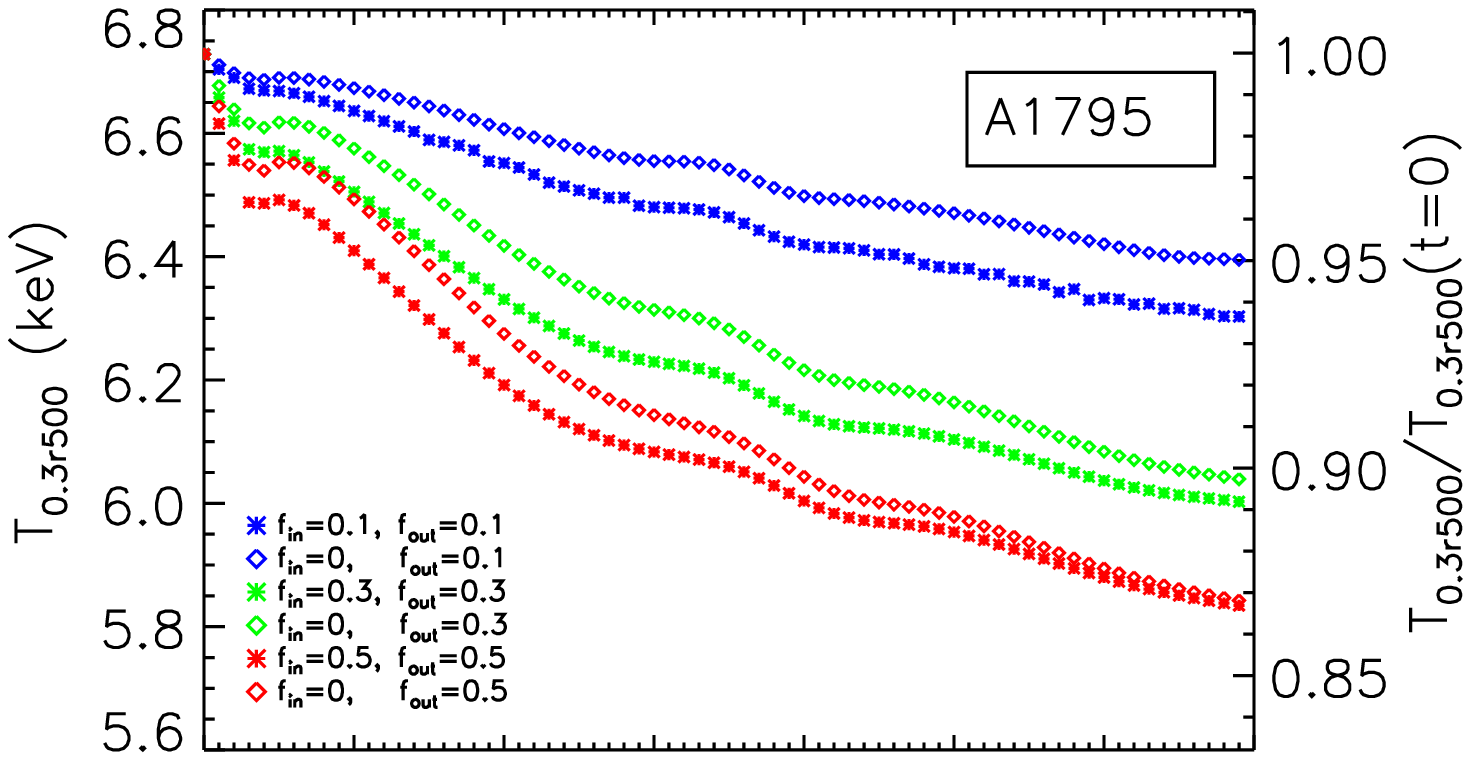}
  \includegraphics[width=0.45\textwidth]{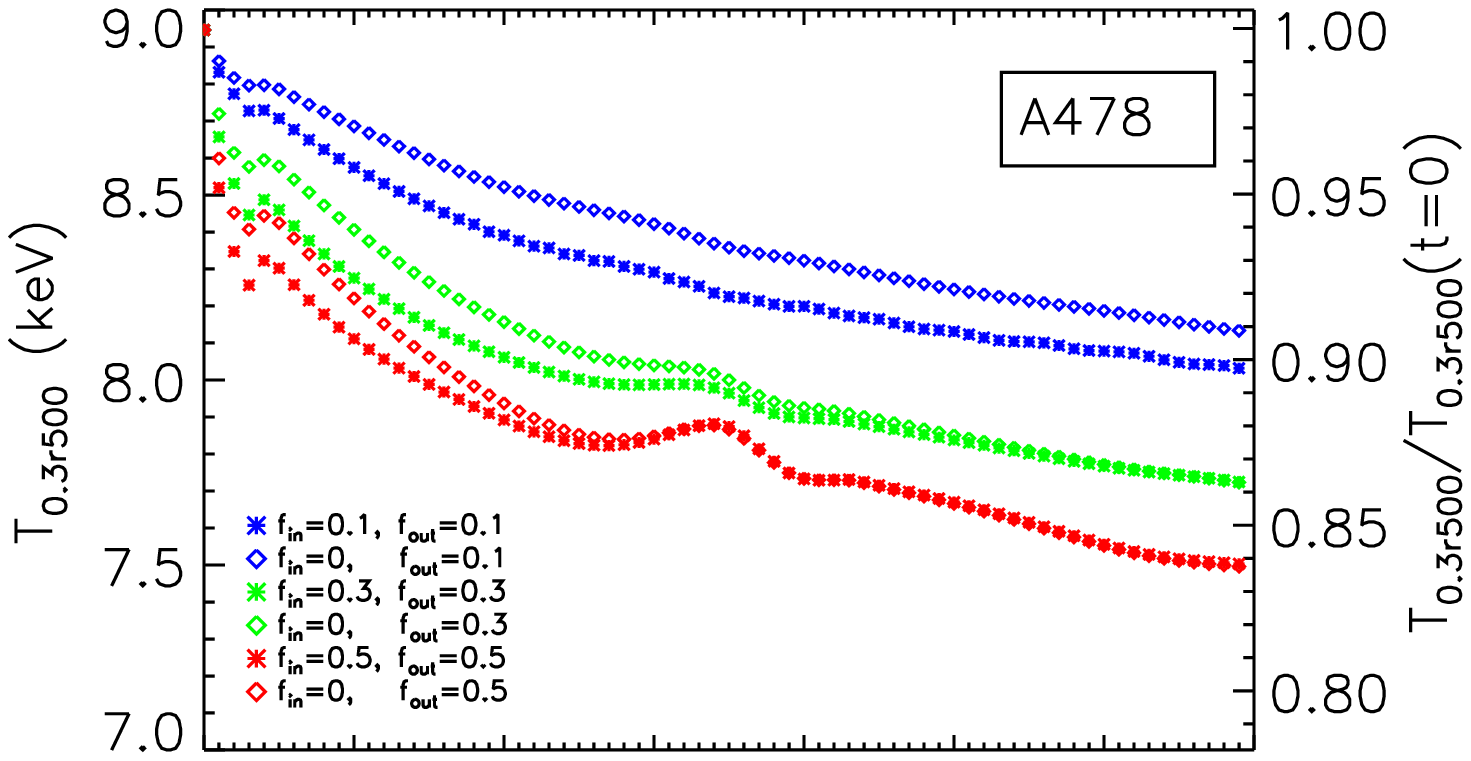}
  \includegraphics[width=0.45\textwidth]{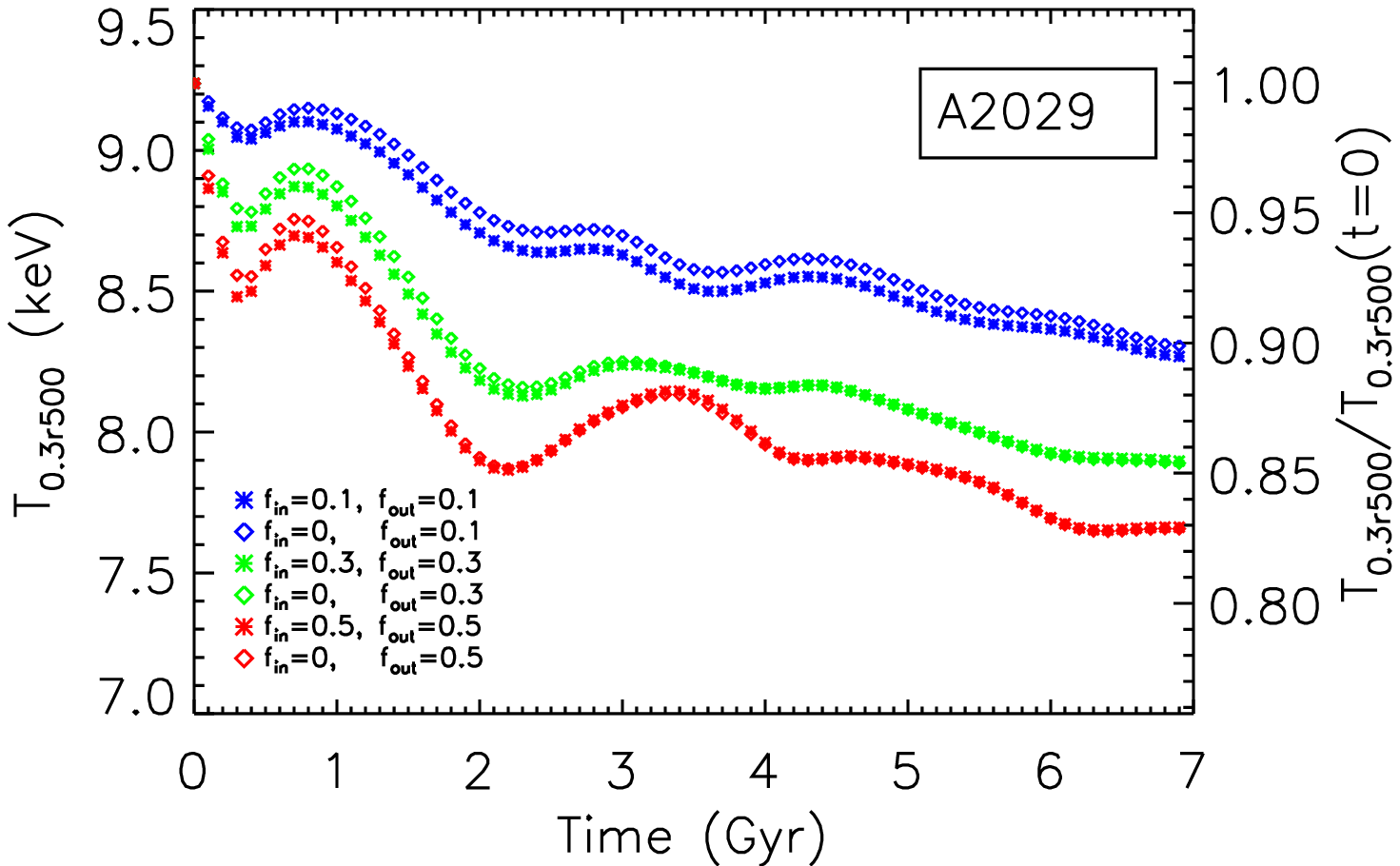}
  \caption{Evolution of the gas temperature at $r=0.3r_{500}$ in our simulations for the cluster A1795 (top), A478 (middle), and A2029 (bottom). Blue, green and red symbols correspond to results in runs with $f_{\rm out}= 0.1$, $0.3$ and $0.5$, respectively. Asterisks denote runs with $f_{\rm in}=f_{\rm out}$, while open diamonds refer to runs with $f_{\rm in}=0$.}
  \label{hightemevo}
\end{figure}

We then investigate how different levels of conductivity affect the temperature evolution in our clusters. Figure \ref{tempevo} shows the evolution of temperature profiles in our simulations for A1795 (top row), A478 (middle row), and A2029 (bottom row) with the conductivity suppression factor $f_{\rm out}=0.1$ (left column), $0.3$ (middle column), and $0.5$ (right column). Figure \ref{tempevo} shows results in simulations with $f_{\rm in}=f_{\rm out}$, while Figure \ref{tempnic} corresponds to runs with $f_{\rm in}=0$ (i.e., with inward conduction prohibited). It is clear that inward thermal conduction from the temperature peak to the cluster center has a dramatic impact on inner cool core regions, especially in runs with $f_{\rm in}=0.3$ or $0.5$, where the core temperature is gradually heated up to be roughly isothermal. This is consistent with previous global stability analyses of cluster models where cool cores could not be stably maintained by thermal conduction alone (\citealt{guo08b}). On the other hand, the temperature evolution in outer regions ($r>r_{\rm peak}$) is not substantially affected by the level of inward conduction in inner regions, as clearly seen by comparing runs with $f_{\rm in}=f_{\rm out}$ shown in Figure \ref{tempevo} to the corresponding ones with $f_{\rm in}=0$ shown in Figure \ref{tempnic} (also see Fig. \ref{hightemevo}). Therefore in the rest of the paper, we limit our analysis to the runs with $f_{\rm in}=f_{\rm out}$.  

As can be seen in the top panels of Figure \ref{tempevo}, the evolution of the outer temperature profile in A1795 depends rather sensitively on the level of thermal conductivity: the larger the conduction suppression factor is, the more dramatic the temperature profile changes. The evolution of the outer temperature profile in run A1Y with $f_{\rm out}=0.1$ may be considered as insignificant, but in runs A3Y and A5Y with $f_{\rm out}=0.3$ and $0.5$, respectively, the temperature evolution is really significant. For higher-temperature systems A478 and A2029 shown in the middle and bottom rows, the evolution of the outer temperature profiles in our runs with $f_{\rm out}=0.1$, $0.3$ or $0.5$ is all quite substantial. Furthermore, Figures \ref{tempevo} and \ref{tempnic} indicate that the temperature evolution in outer regions in these simulations can be roughly separated into two stages: an early fast-evolving stage during $0<t<3$ Gyr, and a later slowly-evolving stage at $t>3$ Gyr. During the former stage in runs with $f_{\rm out}=0.3$ or $0.5$, the outer temperature profile evolves quickly, and particularly gas temperatures between $r_{\rm peak}$ and $\sim 800$ kpc drop substantially within about $3$ Gyr.

In order to quantify the significance of temperature evolution in our simulations, we show in Figure \ref{hightemevo} the evolution of the ICM temperature $T_{0.3r_{500}}$ at $r=0.3r_{500}$ in our simulations for the clusters A1795 (top), A478 (middle), and A2029 (bottom). As shown in Table \ref{table3}, $r=0.3r_{500}>r_{\rm peak}$ is located in outer cluster regions with $dT/dr<0$, and this location in our clusters has also been well studied by {\it Chandra} X-ray observations. Figure \ref{hightemevo} shows clearly that the value of $T_{0.3r_{500}}$ drops with time due to thermal conduction, and the decrease in $T_{0.3r_{500}}$ is more significant in more massive clusters with larger values of $f_{\rm out}$. 
When $f_{\rm out}=0.1$, the value of $T_{0.3r_{500}}$ drops from $t=0$ to $t=3$ Gyr by $\sim 5\%$ in A1795, and by $\sim 10\%$ in the more massive clusters A478 and A2029. When $f_{\rm out}=0.3$ or $0.5$, $T_{0.3r_{500}}$ typically decreases by $\sim 10 - 20\%$ during this fast-evolving stage. Note that the temperature decrease is even more prominent at $r=r_{\rm peak}$.

\begin{figure}
  \includegraphics[width=0.45\textwidth]{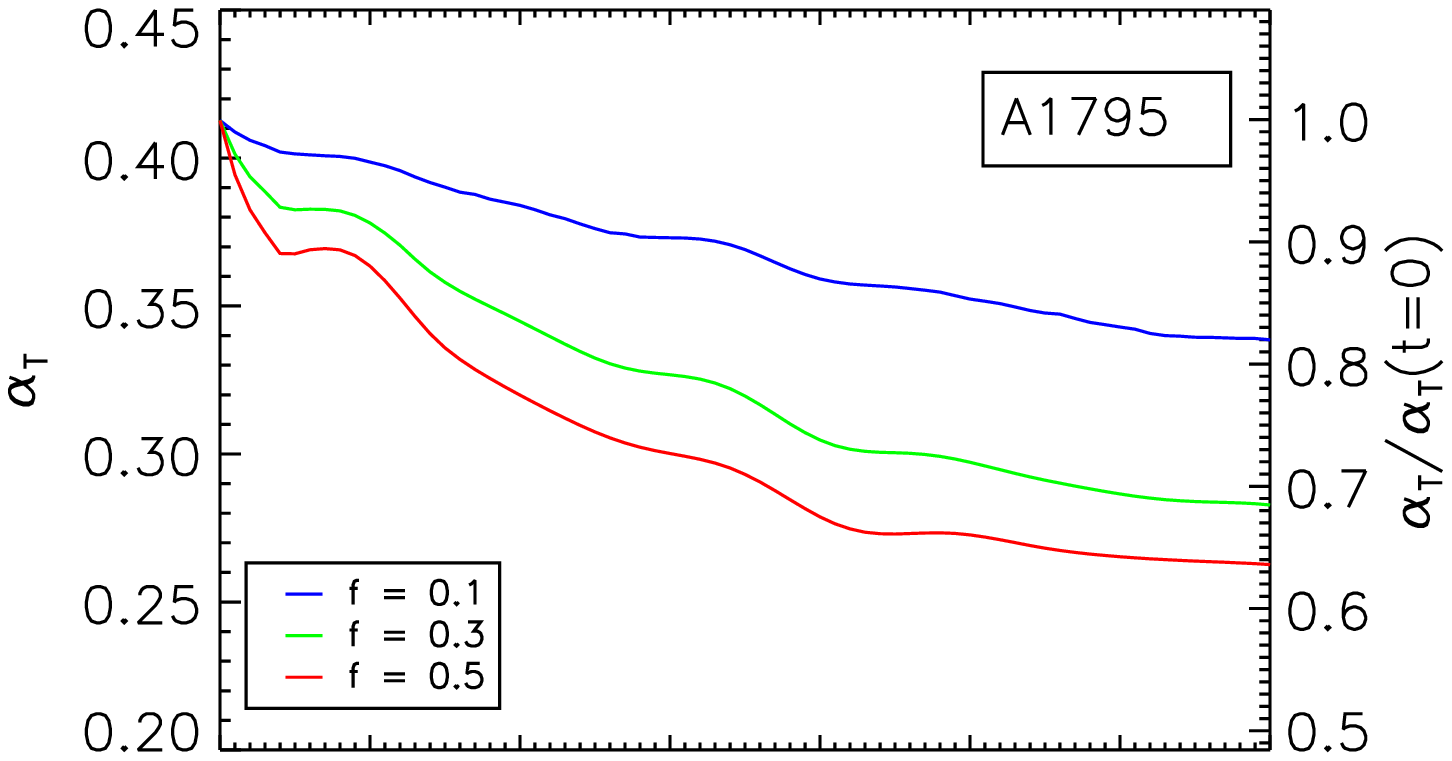}
  \includegraphics[width=0.45\textwidth]{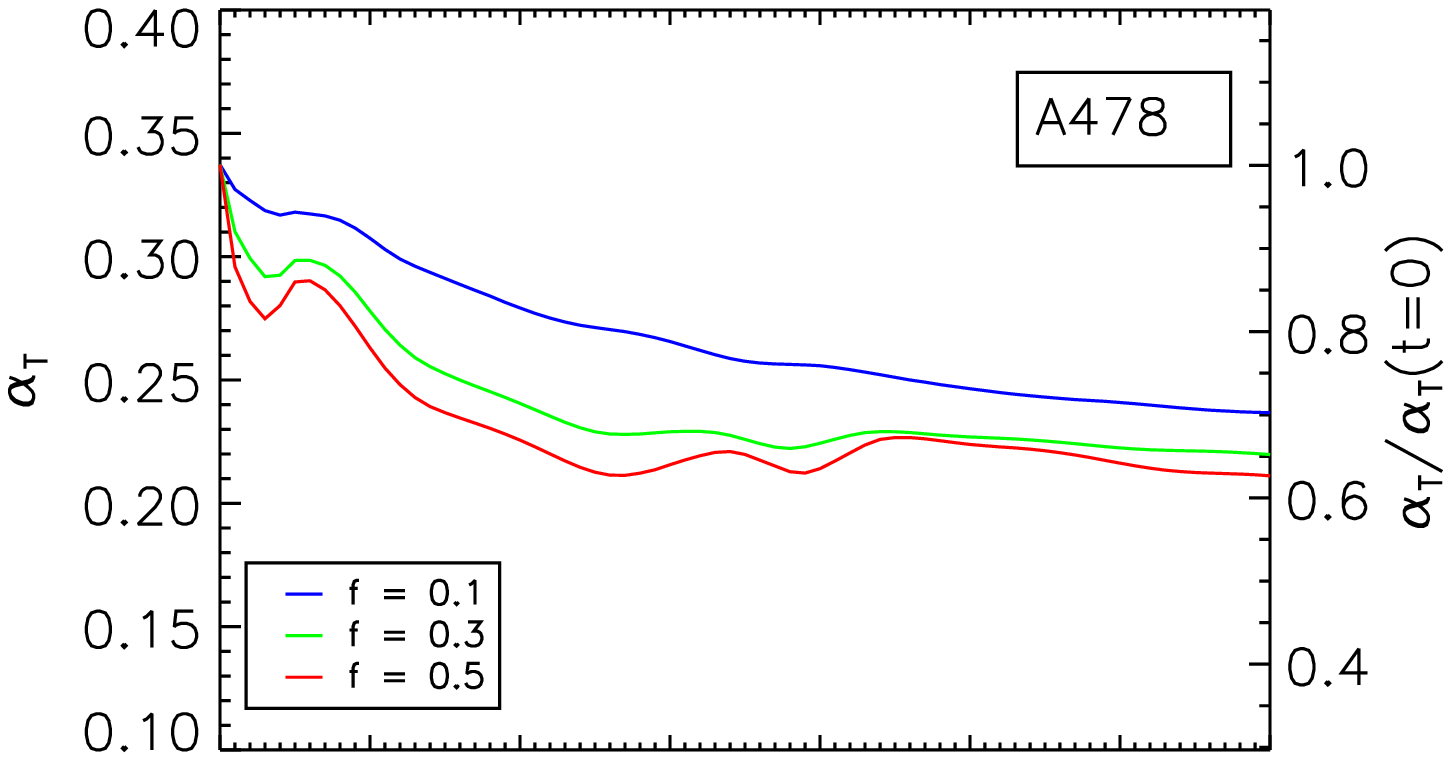}
  \includegraphics[width=0.45\textwidth]{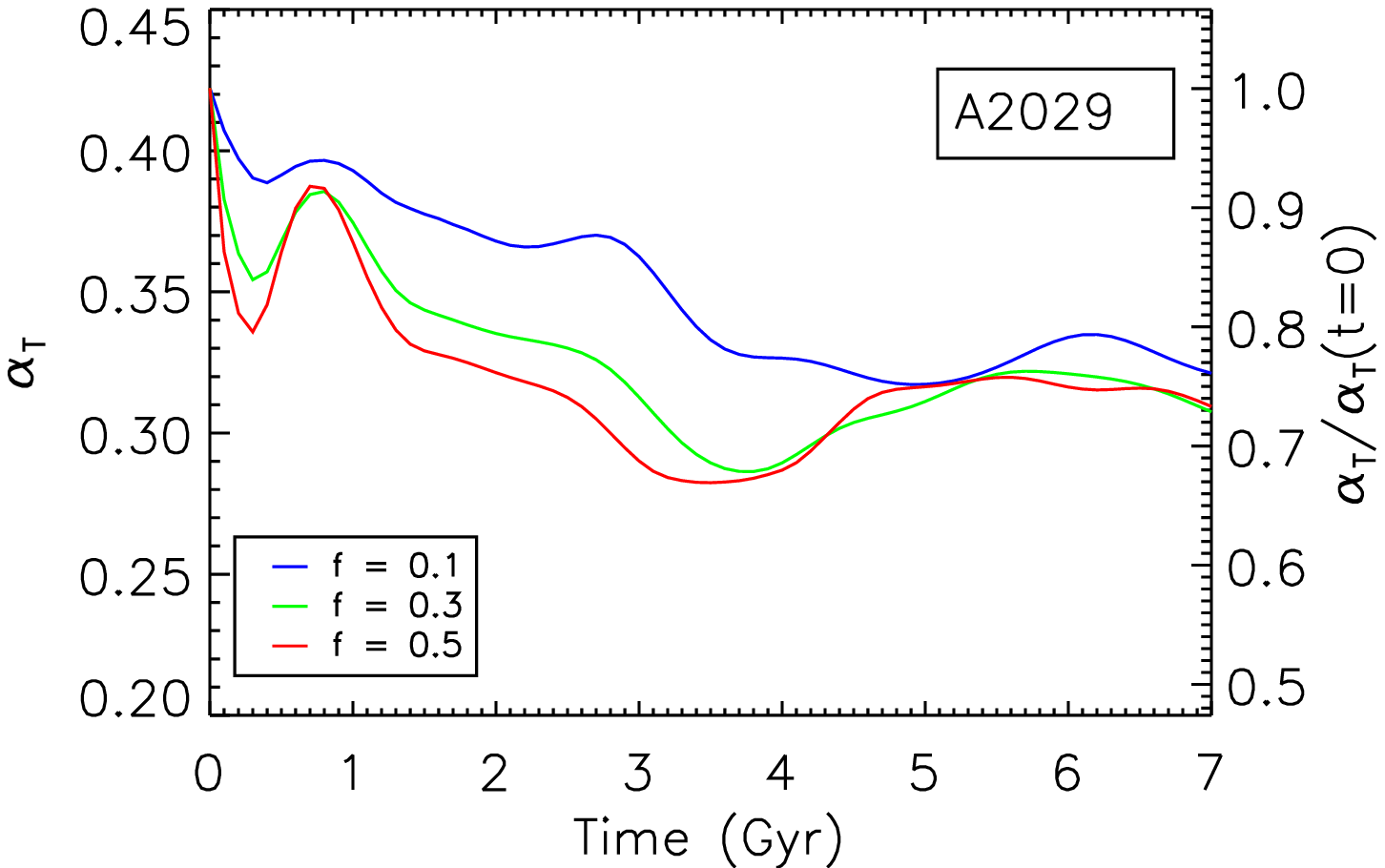}
  \caption{Evolution of the average temperature slope $\alpha_{T}$ between $0.3r_{500}$ and $r_{500}$ in our simulations for the cluster A1795 (top), A478 (middle), and A2029 (bottom). In these simulations, $f_{\rm in}=f_{\rm out}$, and blue, green and red lines show the corresponding results in runs with $f_{\rm out}= 0.1$, $0.3$ and $0.5$, respectively.}
  \label{fig:tslope}
\end{figure}

In addition to the decrease in gas temperatures, the temperature profile in outer cluster regions becomes flatter, as clearly seen in Figures \ref{tempevo} and \ref{tempnic}. To quantify this effect, we calculate the average temperature slope $\alpha_{T}$ between $0.3r_{500}$ and $r_{500}$:
  \begin{equation}
    \alpha_{T} = -\left<\frac{d~ {\rm log}~T}{d ~{\rm log} ~r}\right>=-\frac{{\rm log}~T_{r500}-{\rm log}~T_{0.3r500}}{{\rm log}~r_{500}-{\rm log}~(0.3r_{500})}~{,}   \label{eq:tslope}
 \end{equation}
\noindent
and show its evolution in our simulations for the clusters A1795 (top), A478 (middle), and A2029 (bottom) in Figure \ref{fig:tslope}. It it clear that the drop in the outer temperature slope $\alpha_{T} $ is more significant than that in gas temperatures. When $f_{\rm out}=0.3$ or $0.5$, the value of $\alpha_{T}$ typically decreases by $\sim 30 - 40\%$ during the early fast-evolving stage $0<t<3$ Gyr. Note that for the clusters A478 and A2029, $\alpha_{T}$ drops by $\sim 20\%$ from $t=0$ to $3$ Gyr even when $f_{\rm out}=0.1$.

\begin{figure*}
	\centering
	\begin{center}
		\includegraphics[width=0.3\textwidth]{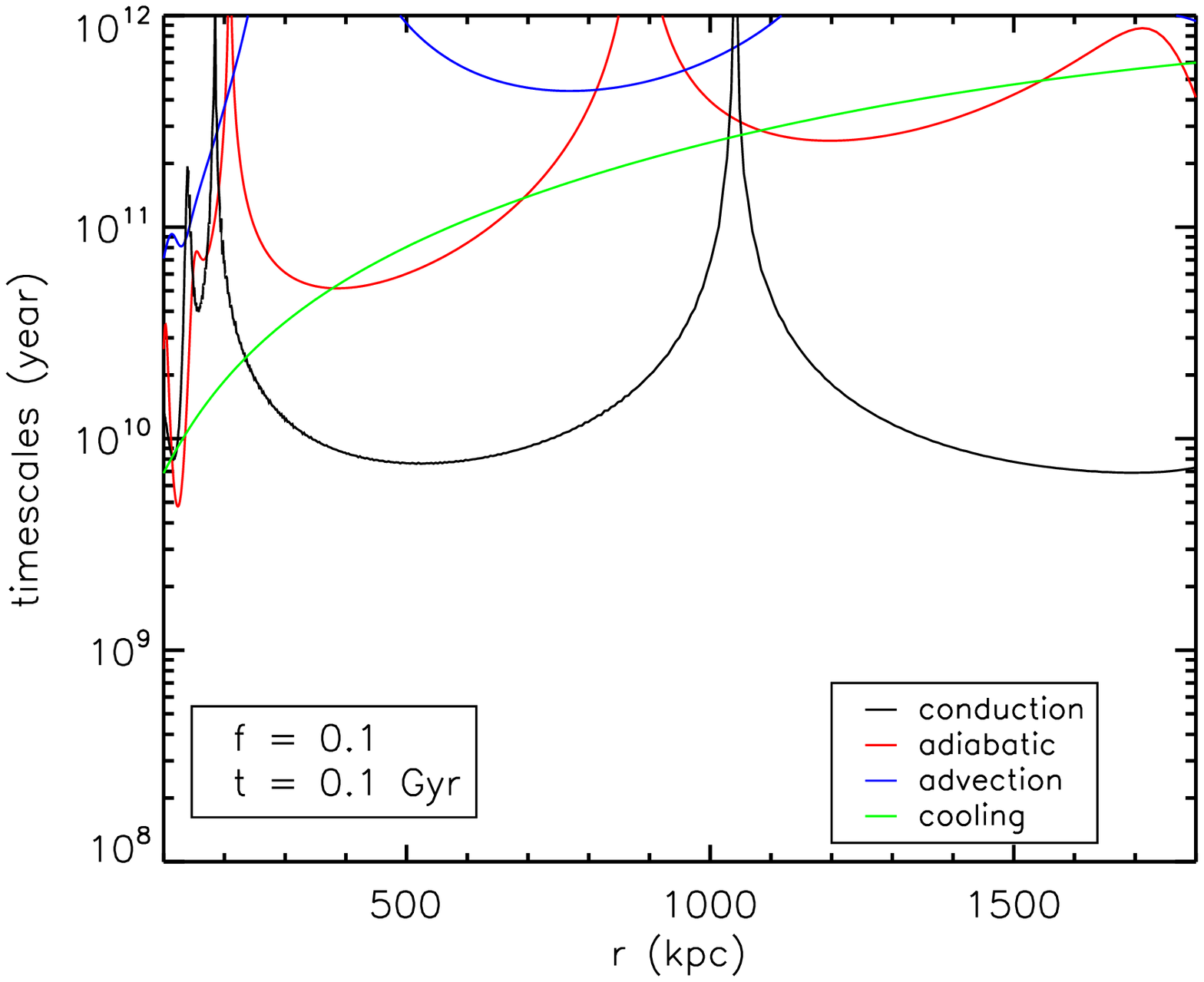}\hspace{0.35cm}
		\includegraphics[width=0.3\textwidth]{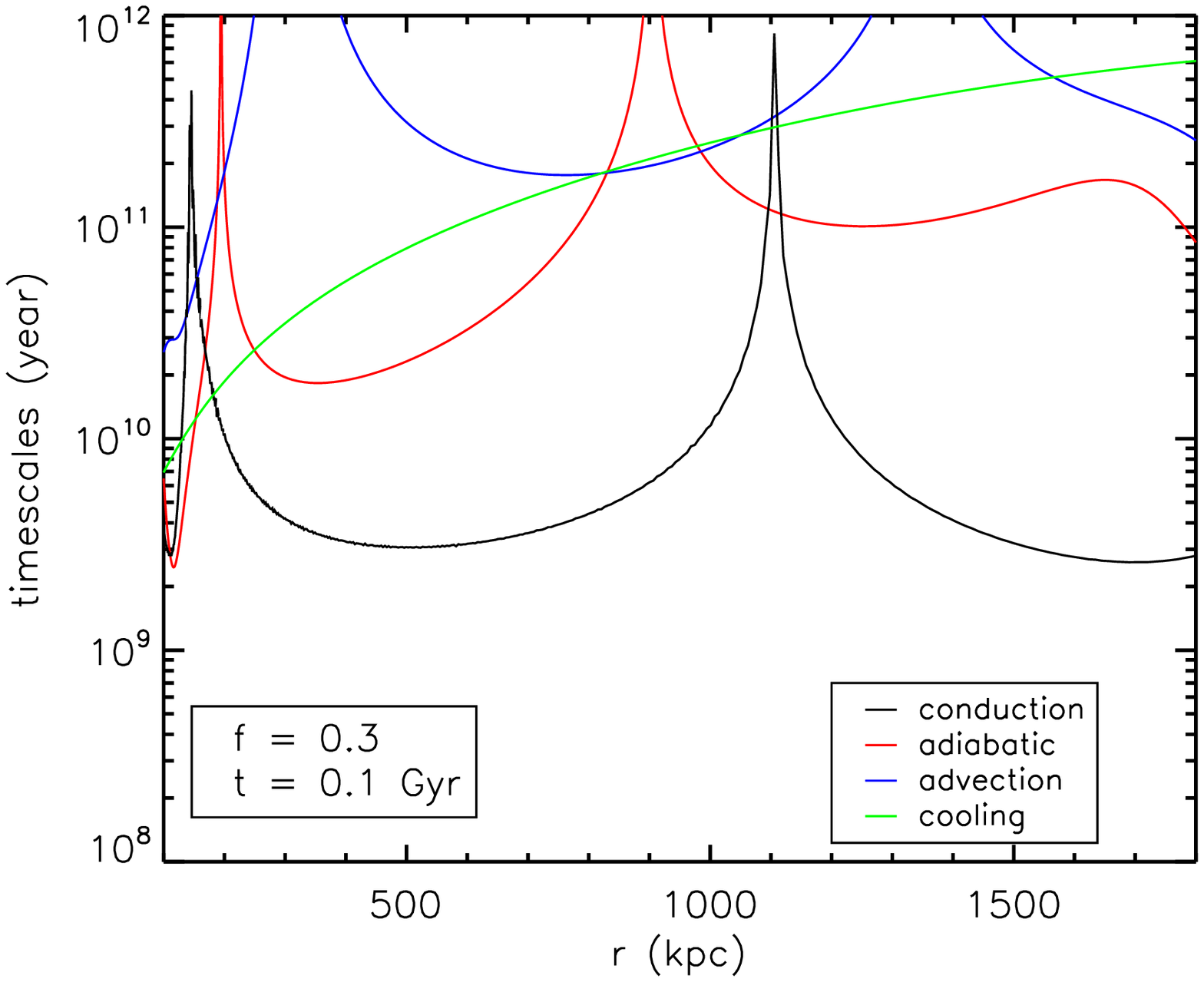}\hspace{0.35cm}
		\includegraphics[width=0.3\textwidth]{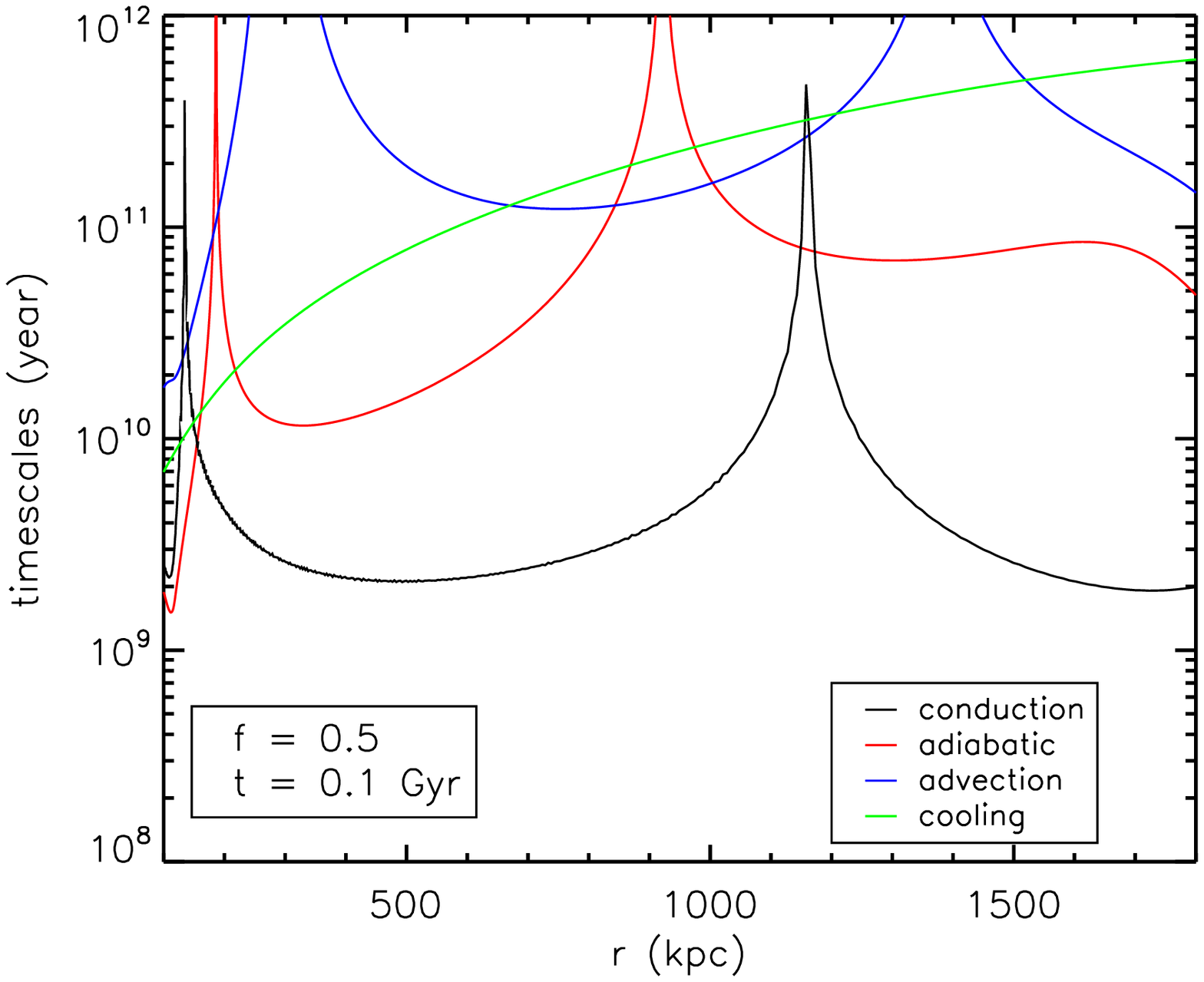}\hspace{0.35cm}
	\end{center}
	\begin{center}
		\includegraphics[width=0.3\textwidth]{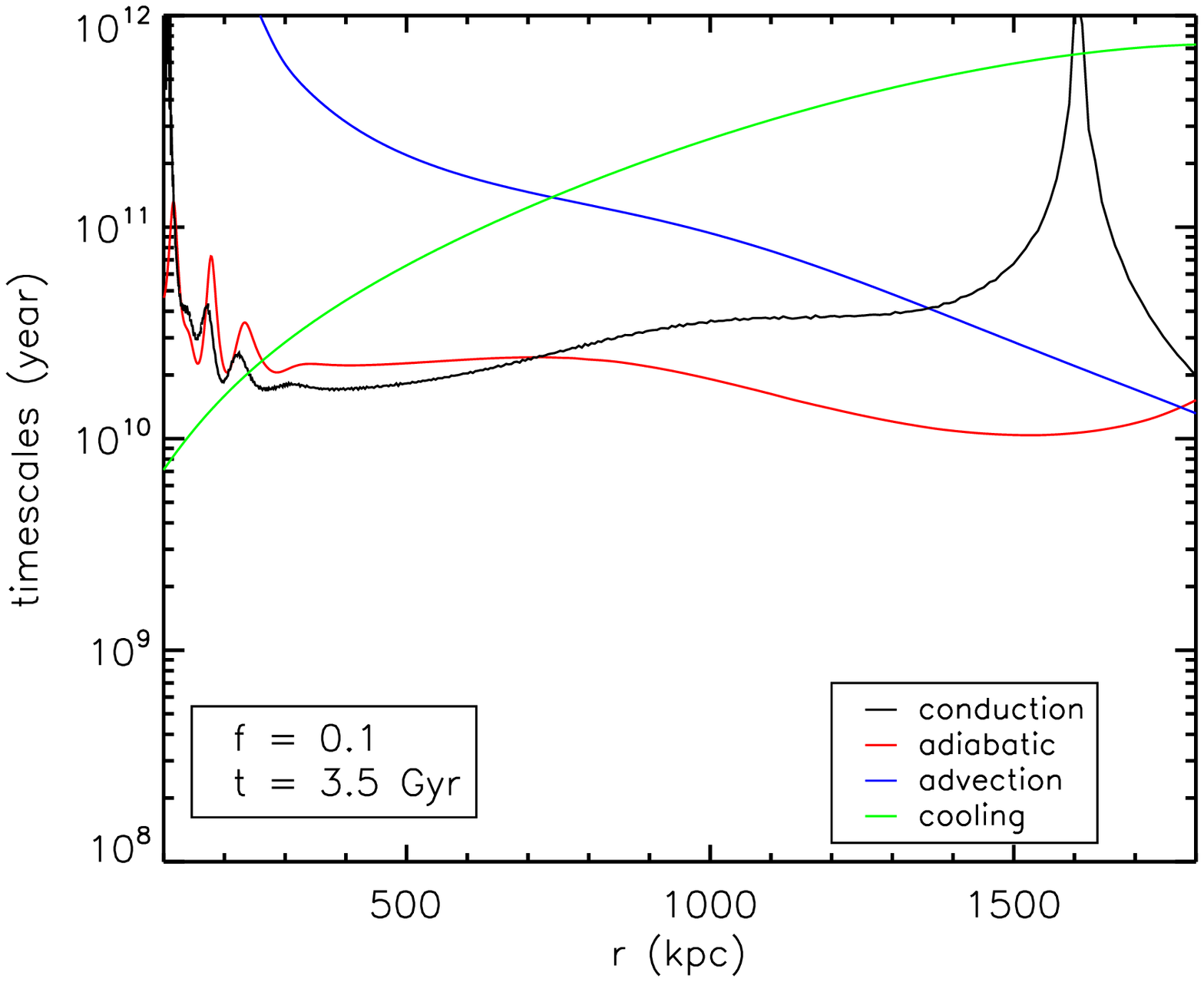}\hspace{0.35cm}
		\includegraphics[width=0.3\textwidth]{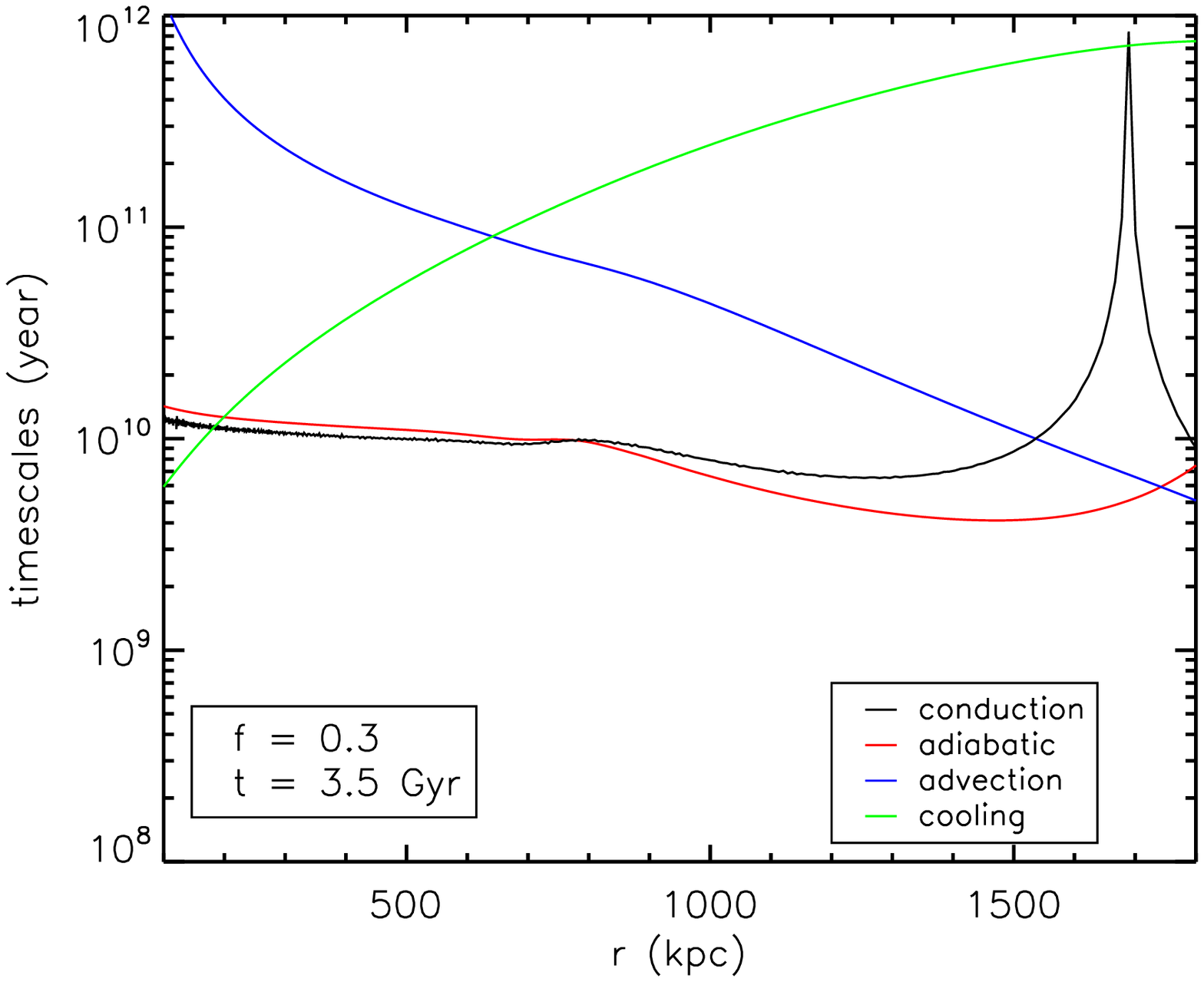}\hspace{0.35cm}
		\includegraphics[width=0.3\textwidth]{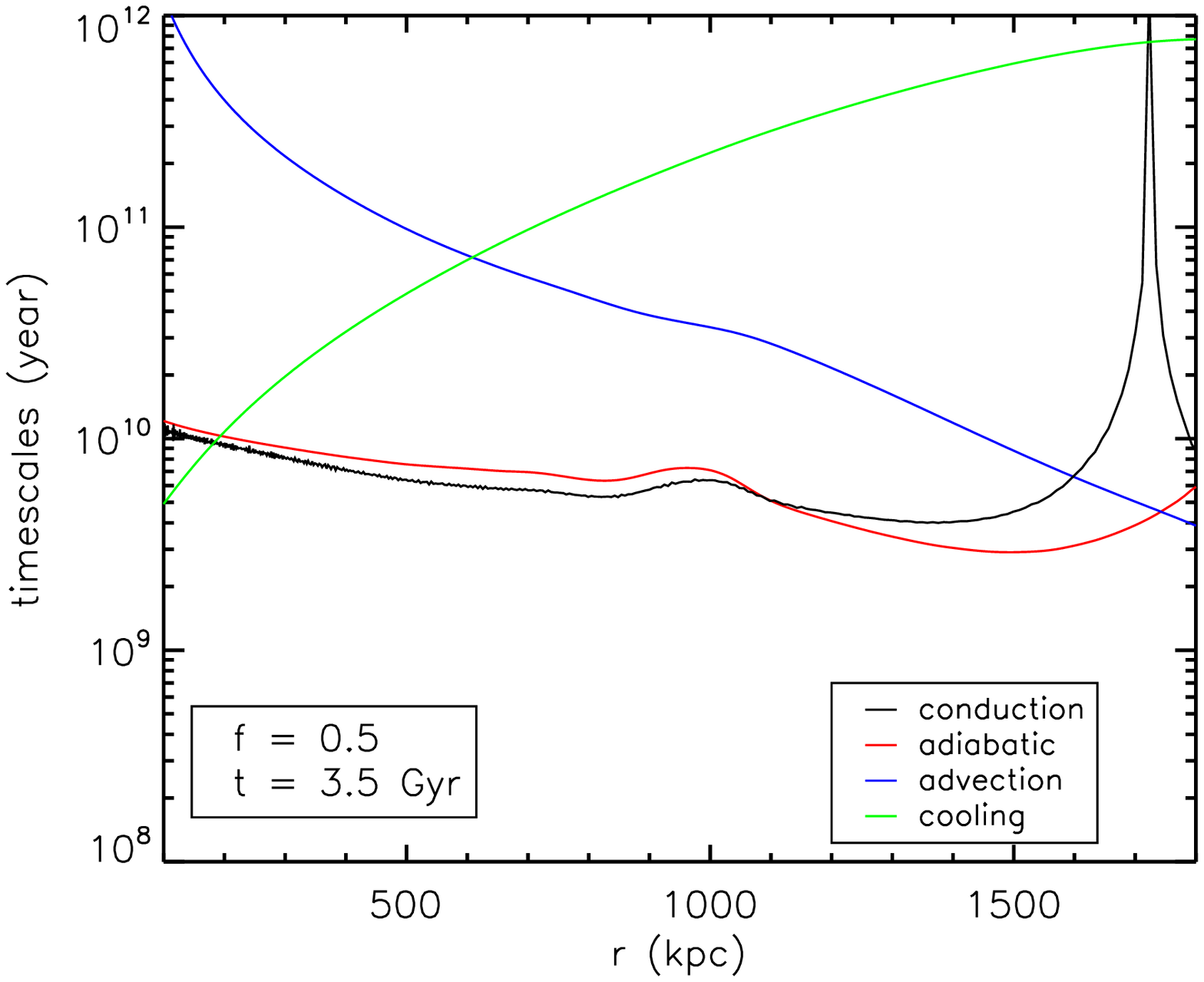}\hspace{0.35cm}
	\end{center}		
	\caption{Temperature variation timescales due to thermal conduction (black), adiabatic compression (or rarefaction; red), and advection (blue) for the cluster A2029 in the simulations with $f=0.1$ (left panels), $0.3$ (middle panels), and $0.5$ (right panels) at $t=0.1$ Gyr (top row) and $3.5$ Gyr (bottom row). Green lines show the local gas cooling time for comparison. In these simulations, $f_{\rm in}=f_{\rm out}=f$.}
	\label{timescales}
\end{figure*}

\subsection{Physics behind the ICM Temperature Evolution}
\label{sec:physics}

What drives the ICM temperature evolution in our simulations? According to the hydrodynamic equations (\ref{hydro1}) - (\ref{hydro3}) and the ideal gas law, the temperature evolution is governed by 
\begin{equation}
\frac{\partial T}{\partial t} = -\frac{T}{e} P \nabla \cdot {\bf v}-\frac{T}{e}\nabla \cdot {\boldsymbol{F}}-{\bf v}\cdot\nabla{T}~{,} 
\label{tempvar}
\end{equation}
where the terms on the right-hand side represent temperature variations due to adiabatic compression (or rarefaction), thermal conduction, and gas advection, respectively. Therefore, we may define the adiabatic timescale
\begin{equation}
  t_{\rm adi} \equiv T/|\frac{T}{e} P\nabla \cdot {\bf v}|  = \frac{1}{(\gamma -1)|\nabla \cdot {\bf v}|}~{,} 
  \label{compresstime}
\end{equation}
conduction timescale
\begin{equation}
  t_{\rm cond} \equiv T/|\frac{T}{e}\nabla \cdot {\boldsymbol{F}}| = \frac{e}{|\nabla \cdot {\boldsymbol{F}}|}~{,} 
\end{equation}
and advection timescale
\begin{equation}
  t_{\rm adv} \equiv T/|{\bf v}\cdot\nabla{T}|~{,} 
\end{equation}
for the temperature evolution as a function of radius in galaxy clusters.

Figure \ref{timescales} shows the above timescales for the cluster A2029 at $t=0.1$ Gyr (top row) and $3.5$ Gyr (bottom row) in runs C1Y (left column; $f=0.1$), C3Y (middle column; $f=0.3$), and C5Y (right column; $f=0.5$). For comparison, also shown is the gas cooling time (green lines), defined as $t_{\rm cool} = e / (n_{\rm e}n_{\rm i}\Lambda)$, where $n_{\rm e}n_{\rm i}\Lambda(T)$ is the gas cooling rate per unit volume adopted from equation (35) of \citet{guo08}, which is an analytic expression originally given by \citet{tozzi01} to fit cooling rates in \citet{SutherlandDopita1993} at metallicity $Z = 0.3 Z_{\odot}$. From Figure \ref{timescales}, it is clear that the gas cooling time in outer regions $r>r_{\rm peak}$ ($r_{\rm peak}=329$ kpc for A2029) is longer than $10$ Gyr, which justifies our neglect of radiative cooling in our simulations. Note that radiative cooling is important in central regions ($r\lesssim 100$ kpc), where AGN feedback may play a key role in offsetting cooling (e.g., \citealt{guo08}; \citealt{guo18}).

The top row of Figure \ref{timescales} clearly shows that, at $t=0.1$ Gyr, the conduction timescale $t_{\rm cond}$ in most outer regions is significantly shorter than all other timescales in all of our simulations with $f=0.1$, $0.3$, and $0.5$, indicating that the early temperature evolution in outer regions is dominated by thermal conduction. The conduction timescale $t_{\rm cond}$ also decreases as the value of $f$ increases, and when $f=0.3$, or $0.5$, $t_{\rm cond} \lesssim 4$ Gyr in most outer regions. This confirms that the early fast temperature evolution during $0<t<3$ Gyr discussed in Section \ref{sec:Tevol} is driven by thermal conduction. Note that thermal conduction operates as a cooling mechanism for the ICM between the two $t_{\rm cond}$ peaks located at $r\sim 150$ and $1100$ kpc, respectively. For the very inner and outer regions, conduction acts as a heating source for the ICM. These two $t_{\rm cond}$ peaks, which refer to the locations with $\nabla \cdot {\boldsymbol{F}}=0$, move with time, and the region with conductive cooling expands during our simulations.

The bottom panels of Figure \ref{timescales} show that at $t=3.5$ Gyr, the adiabatic timescale becomes comparable to the conduction timescale between $\sim 100$ kpc and $r_{500}$, indicating that compressional heating is competing with conductive cooling in these regions. Both timescales at this time are longer than $4$ Gyr in most regions between $100$ and $1000$ kpc, which is consistent with the later slowly-evolving stage at $t>3$ Gyr found in Section \ref{sec:Tevol}.

The results for the other two clusters A1795 and A478 are very similar, except that the conduction timescale in A1795 is slightly longer due to its relatively lower temperature and thus conductivity. X-ray observations show that, when scaled with the average cluster temperature and the virial radius, the temperature profiles in outer regions ($r \gtrsim 0.1r_{200}$) of massive clusters are remarkably similar (e.g., \citealt{Vikhlinin2005}; \citealt{pratt07}), suggesting that our results in Sections \ref{sec:Tevol} and \ref{sec:implications} are very robust.

\subsection{Implications and Discussions}
\label{sec:implications}

Let us summarize our main findings in Sections \ref{sec:Tevol} and \ref{sec:physics}. When $f=0.3$ or larger, the ICM in outer regions of massive clusters would not stay in a long-term equilibrium state due to the loss of thermal equilibrium, and thermal conduction along the radial direction would lead to a decrease of the ICM temperature by $\sim 10 - 20\%$ within $3$ Gyr at a representative radius of $r=0.3r_{500}$. During the same period, the average temperature slope $\alpha_{T}$ between $0.3r_{500}$ and $r_{500}$ typically decreases by $\sim 30 - 40\%$. Even when $f=0.1$, $T_{0.3r_{500}}$ drops by $\sim 10\%$ and $\alpha_{T}$ drops by $\sim 20\%$ within 3 Gyr for massive clusters such as A478 and A2029. 
 
However, when scaled with the average cluster temperature and the estimated virial radius, the observed temperature profiles in outer regions of galaxy clusters show remarkable similarity (\citealt{Vikhlinin2005}; \citealt{pratt07}). Although these observed profiles still have substantial (typically $\sim10\%$-$20\%$) scatters around the average temperature profile, it is notable that the same average profile applies to clusters covering a substantial range in mass, including very massive clusters such as A478 and A2029 with relatively short conduction timescales, suggesting that this universal temperature distribution may be long-lasting and stable. If this is indeed the case, our simulations indicate that the effective thermal conductivity along the radial direction in outer regions of massive clusters must be suppressed below the classic Spitzer value by a factor of $10$ or more (i.e., $f\lesssim 0.1$).

Thermal conduction along magnetic field lines may be strongly suppressed by kinetic mirror or whistler instabilities on the length scale of the ion or electron gyroradius (\citealt{Komarov2016}; \citealt{Riquelme16}; \citealt{clark16}). \citet{Komarov2016} show that mirror instability alone could suppress conductivity along field lines by a factor of $5$. Assuming that radial conduction is further suppressed by another factor of $5$ due to tangled magnetic fields in turbulence \citep{NarayanMedvedev2001}, radial conductivity is roughly $\sim 1/25$ of the Spitzer value, consistent with our current results. 

On the other hand, if thermal conduction along magnetic field lines is not suppressed, the MTI may develop in outer regions of clusters with negative temperature gradients. When it saturates, magnetic field lines become largely radial, leading to efficient thermal conduction along the radial direction with $f\sim 0.5$ \citep{Parrish2008}, which is clearly in contradiction with the observed stable temperature profiles in outer regions of massive clusters. However, the development of MTI may be affected by g-mode overstabilities \citep{Balbus2010} or external turbulence driven by cosmic accretion (e.g. merging substructures), which may further lead to tangled magnetic field lines. Due to strong suppression of effective conductivity by field-line tangling \citep{RechesterRosenbluth1978}, this argument may potentially hinder us to put a strong constraint on the efficiency of conduction along magnetic field lines. While a comprehensive analysis is beyond the scope of the present paper, we note that if turbulence cascades and extends over a wide range of length scales, the effective conductivity may reach $\sim1/5$ of the Spitzer value \citep{NarayanMedvedev2001}, which is still too large to be consistent with observations.

If true, the suppression of parallel conduction along magnetic field lines would have profound implications in several important topics in astrophysics. It could easily explain the existence of sharp temperature jumps across cold fronts observed in many galaxy clusters \citep{zuhone16}. It could affect the slope of the power spectrum of gas density perturbations in galaxy clusters (\citealt{gaspari13}; \citealt{eckert17}) and explain the long-lasting temperature gradients observed in ram-pressure-stripped gas tails in galaxy clusters \citep{grandi16}. It implies that thermal conduction contributes negligibly to heating cool cores of galaxy clusters, corroborating the current paradigm that AGN feedback plays the key role in solving the cooling flow problem \citep{mcnamara12}. It also has important implications for hot circumgalactic medium around galaxies (e.g., the Milky Way; \citealt{guo12}; \citealt{fang13}) and the local thermal instability therein \citep{tumlinson17}, and for hot accretion flows around black holes (\citealt{bu11}; \citealt{yuan14}).  

If radial conductivity in real clusters is larger than $\sim 1/10$ of the Spitzer value (i.e. $f\gtrsim 0.1$), our study implies that additional heating sources must operate in outer regions of massive clusters to offset the effect of thermal conduction and maintain the observed temperature distributions. One potential heating source is accretion shocks formed when cosmic baryonic matter falls into galaxy clusters (e.g., \citealt{mccourt13}). However, cosmological simulations indicate that for relaxed galaxy clusters, dark matter haloes are typically in the later slow accretion phase when cosmic accretion mainly affects very outer regions and the mean radial velocity of dark matter within $r\lesssim r_{200}$ is about zero (\citealt{zhao03}; \citealt{cuesta08}). This suggests that cosmic accretion may not have a strong impact on the ICM regions of our interest (say, $r\lesssim r_{500}$). Furthermore, the observed temperature profiles in the outer regions of massive clusters are roughly consistent with cosmological hydrodynamic simulations where thermal conduction is not considered (e.g., \citealt{Loken2002}; \citealt{Borgani2004}; \citealt{hahn17}). Future cosmological simulations incorporating thermal conduction would be helpful to investigate this issue in more details.

\section{Summary}
\label{section:summary}

As one major type of low-collisionality plasma in the Universe, the hot, diffuse ICM is the dominant baryonic component in galaxy clusters. Due to the high temperatures of the ICM, thermal conduction is expected to be very efficient along magnetic field lines in the traditional kinetic theory \citep{Spitzer1962}. While complicated by the existence of turbulence and magnetic fields, the efficiency and effects of thermal conduction in the ICM have thus received a lot of attention in the literature. The outer regions of galaxy clusters typically have radially-declining temperature profiles, and previous studies suggest that the MTI may develop in these regions and quickly reorient magnetic field lines to be largely radial, resulting in efficient thermal conduction along the radial direction (\citealt{balbus00}; \citealt{ParrishStone2007}; \citealt{Parrish2008}; \citealt{Sharma2008}). Such efficient radial conduction is expected to have a substantial impact on the observed temperature profiles in the outer cluster regions, which may potentially be an ideal testbed to investigate the efficiency of thermal conduction in low-collisionality plasmas.

In this paper, we performed a series of hydrodynamic simulations to investigate the impact of thermal conduction on the radially-declining temperature profiles in outer regions of three representative massive clusters, A1795, A478 and A2029, which have all been well observed by {\it Chandra} X-ray observations up to a relatively large radius of nearly $r_{500}$ or beyond. Starting with the observed temperature and density distributions in hydrostatic equilibrium, we investigate how different levels of thermal conduction affect the evolution of the ICM temperature profile. Our simulations clearly indicate that thermal conduction substantially modifies the ICM temperature profile, whose evolution can be roughly separated into two stages: an early fast-evolving stage during $0<t<3$ Gyr dominated by thermal conduction, and a later slowly-evolving stage at $t>3$ Gyr under the effects of both conduction and adiabatic compression. When the conductivity suppression factor $f$ is $0.3$ or larger, the gas temperature at a representative radius of $0.3r_{500}$ typically decreases by $\sim 10 - 20\%$ and the average temperature slope $\alpha_{T}$ between $0.3r_{500}$ and $r_{500}$ drops by $\sim 30 - 40\%$ during the first stage. Even when $f=0.1$, $T_{0.3r_{500}}$ drops by $\sim 10\%$ and $\alpha_{T}$ drops by $\sim 20\%$ within 3 Gyr for massive clusters such as A478 and A2029. 

Our simulations therefore indicate that the outer regions of massive clusters would not remain in a long-term equilibrium state if the effective thermal conductivity along the radial direction is larger than $\sim 1/10$ of the classic Spitzer value. However, X-ray observations show that the outer regions of massive clusters have remarkably similar radially-declining temperature profiles (e.g., \citealt{Vikhlinin2005}; \citealt{pratt07}), suggesting that the observed temperature distributions in outer cluster regions are usually long-lasting and stable. Our study thus suggests that the effective conductivity along the radial direction in outer cluster regions must be suppressed below the Spitzer value by a factor of $10$ or more, unless additional heating sources offset conductive cooling, maintaining the observed temperature distributions. 

Our results are consistent with recent theoretical studies on the suppression of parallel conduction along magnetic field lines in low-collisionality plasmas by mirror or whistler instabilities on kinetic scales (\citealt{Komarov2016}; \citealt{Riquelme16}; \citealt{clark16}). On the other hand, if thermal conduction along magnetic field lines operates unimpeded, effective conduction along the radial direction is expected to be quite efficient in outer cluster regions, no matter whether the MTI or external turbulence dominates ($f\sim 0.5$ and $\sim 0.2$ respectively). This is in contradiction with the rather stable outer temperature profiles of massive clusters commonly observed by X-ray observations. Our study may provide a smoking-gun evidence for the suppression of parallel conduction along magnetic field lines by kinetic mirror or whistler instabilities in low-collisionality plasmas.

\acknowledgements 

We are grateful to Eugene Churazov, Sergey Komarov, and an anonymous referee for insightful comments and suggestions. This work was supported partially by Natural Science Foundation of China (Grant No. 11633006, 11725312, 11643001, U1431228, 11233003, 11421303, and 11703022), Natural Science Foundation of Shanghai (No. 18ZR1447100), and Chinese Academy of Sciences through the Hundred Talents Program and the Key Research Program of Frontier Sciences (No. QYZDB-SSW-SYS033 and QYZDJ-SSW-SYS008). GM acknowledges the support by the Fundamental Research Funds for the Central Universities (WK2030220017). XEF thanks Zhen-Yi Cai for the help on scientific plotting. Some simulations presented in this work were performed using the high performance computing resources in the Core Facility for Advanced Research Computing at Shanghai Astronomical Observatory.

\bibliography{ms}

\end{document}